\documentclass[twocolumn]{aastex631}
\usepackage{natbib, color, subfigure,amsmath, soul, verbatim}
\usepackage[T1]{fontenc}
\newcommand{\hbeta}{H{$\beta$}\,}

\def\FeII{Fe\,{\tt II}\,}

\defcitealias{Kubota&Done2018}{KD18}
\defcitealias{Kubota&Done2019}{KD19}
\defcitealias{WuQ+2025}{W25}

\def\Lopt{\textit{$\lambda L_{5100}$}}
\def\kms{\textit{$\rm km\,s^{-1}$}}
\def\keV{\textit{$\rm keV$}}

\def\ergs{\textit{$\rm erg\ s^{-1}$}}
\def\Mbh{\textit{$M_{\rm BH}$}}
\def\msun{\textit{$M_{\odot}$}}
\def\LLEdd{\textit{$L/L_{\rm Edd}$}}
\def\LobsLEdd{\textit{$L_{\rm bol, obs}/L_{\rm Edd}$}}
\def\dotm{\textit{$\dot{m}$}}

\def\QH{\textit{$Q_{\rm H}$}}
\def\PhiH{\textit{$\Phi_{\rm H}$}}
\def\nH{\textit{$n_{\rm H}$}}
\def\Rg{\textit{$R_{\rm g}$}}

\shorttitle{BLR photoionization in slim disks}
\shortauthors{Wu et~al.}
\graphicspath{./} 
\begin{document}

\title{Understanding the Broad-line Region of Active Galactic Nuclei with Photoionization. II. Slim disks, Self-shadowing, and BLR sizes} 

\author[0000-0003-4202-1232]{Qiaoya Wu}
\email{qiaoyaw2@illinois.edu}
\affiliation{Department of Astronomy, University of Illinois at Urbana-Champaign, Urbana, IL 61801, USA}

\author[0000-0003-1659-7035]{Yue Shen}
\email{shenyue@illinois.edu}
\affiliation{Department of Astronomy, University of Illinois at Urbana-Champaign, Urbana, IL 61801, USA}
\affiliation{National Center for Supercomputing Applications, University of Illinois at Urbana-Champaign, Urbana, IL 61801, USA}

\author[0000-0002-1065-7239]{Chris Done}
\affiliation{Centre for Extragalactic Astronomy, Department of Physics, Durham University, South Road, Durham DH1 3LE, UK}

\author[0000-0002-2908-7360]{Michael R. Goad}
\affiliation{Department of Physics and Astronomy, University of Leicester, University Road, Leicester LE1 7RH, UK}

\author[0000-0002-5075-7920]{Scott Hagen}
\affiliation{IFPU - Institute for Fundamental Physics of the Universe, Via Beirut 2, 34151 Trieste, Italy}
\affiliation{SISSA - International School for Advanced Studies, Via Bonomea 265, 34136 Trieste, Italy}
\affiliation{INAF - Osservatorio Astronomico di Trieste, Via G. B. Tiepolo 11, I-34143 Trieste, Italy}

\begin{abstract}
Reverberation-mapping (RM) measurements have revealed that high-accretion-rate active galactic nuclei (AGNs) systematically lie below the canonical broad-line region (BLR) radius – optical continuum luminosity ($R-L$) relation, exhibiting shorter lags than those predicted for fixed 5100\,\AA\ luminosity. The physical origin of these offsets remains debated. 
We investigate how accretion-flow structure and BLR cloud properties affect the emissivity-weighted BLR radius using analytic slim-disk SEDs and photoionization calculations on a two-dimensional axisymmetric spatial grid. As the accretion rate approaches and exceeds the Eddington limit, geometric thickening of the inner disk produces anisotropic illumination and self-shadowing, reducing the ionizing flux seen by low-latitude BLR clouds and flattening the $R-L$ relation at high \LLEdd. Self-shadowing at high accretion rates reproduces the observed $R-L$ trend in the RM AGN sample reasonably well, but this effect alone is insufficient to explain the observed lag offset in low-\Mbh\ ($\sim 10^{7}\,M_\odot$) systems with high accretion rates. 
Motivated by accretion-disk density scalings, we further explore models in which the BLR gas density increases toward lower black hole mass or higher accretion rate. 
We find that an accretion-rate-dependent BLR density enhancement further improves the agreement with observed RM data, where the BLR gas density increases by a factor of $3-5$ for one dex increase in $\dotm$. 
Variations in BLR opening angles produce a less important effect on BLR sizes. These results demonstrate that self-consistent modeling of the accretion disk SED, BLR illumination and photoionization, and gas density variations can fully explain the observed distribution of AGNs in the BLR size - optical luminosity plane. This framework provides a physically motivated link between accretion-flow structure and BLR observables across a broad range of black-hole properties.
\end{abstract}

\keywords{black hole physics --- galaxies: active --- quasars}

\section{Introduction}\label{sec:intro}

Active galactic nuclei (AGN) are powered by the accretion of matter onto supermassive black holes (SMBHs) at the centers of massive galaxies. The primary observational features of these systems are well-characterized by accretion disk and photoionization models. The standard thin accretion disk model \citep[SSD;][]{Shakura&Sunyaev1973, Novikov&Thorne1973, Laor&Netzer_1989} has been successfully applied to objects with moderate Eddington ratios $\LLEdd\lesssim0.3$. As the accretion rate increases toward the Eddington limit $\LLEdd\sim1$,  radiation pressure becomes dynamically important, and photon trapping makes advective cooling significant, driving the transition from an SSD to a slim disk \citep{Abramowicz1988}. In this regime, the accretion flow is characterized by a geometrically thick inner region in which the disk height becomes comparable to the radius. Slim disk models have been extensively studied analytically to interpret the observational properties of super-Eddington accreting black holes (BHs) \citep[e.g.][]{Watarai+2000, Mineshige+2000, Chen&Wang2004, Wang2014_selfshadow}. 

The central accretion flow serves as the primary ionizing source for the surrounding environment, providing the radiation field that gives rise to the broad emission lines (BELs) characteristic of AGN UV-optical spectra \citep{Woltjer_1959}. The gas within the broad-line region (BLR) mostly follows virial motion, orbiting within the gravitational potential of the central SMBH at velocities of several thousand \kms. Consequently, the widths of these spectral lines encode vital information regarding both the BH mass and the structure of the BLR \citep{Peterson1993, Peterson2004, Pancoast_2014_lick}. Over the past several decades, reverberation mapping (RM) has been the primary method for measuring the physical size of the BLR. By combining the measured BLR time lag ($\tau_{\rm{BLR}}$) from RM with the virial velocity inferred from the BEL width, the SMBH mass can be determined via the virial assumption: $\Mbh = f\,c\,\tau_{\rm{BLR}}\,\Delta V^2/G$, where $f$ is a dimensionless virial coefficient governing the BLR geometry, dynamics, and viewing angle, c is the speed of light, $\tau_{\rm BLR}$ is the RM lag, $\Delta V$ is the line-width measure of the BEL, and $G$ is the gravitational constant \citep{Peterson2004}.
Decades of RM campaigns have revealed a tight correlation between the optical continuum luminosity measured at $5100$~\AA\ and the average size of the H$\beta$-emitting BLR \citep{Kaspi2000, Bentz2013}. This radius--luminosity ($R-L$) serves as the foundation for single-epoch virial mass estimators \citep{Vestergaard&Peterson2006, Shen2013}. These single-epoch estimators are widely adopted to determine BH masses in high-redshift quasars \citep[e.g.][]{Willott_etal2010, WangF_etal2021} and large spectroscopic surveys \citep{Shen+2011, Wu&Shen2022}, underpinning modern studies of quasar demographics and BH evolution across the universe.

While the $R-L$ relation is relatively tight for sub-Eddington AGNs, recent RM studies targeting sources across a broader range of luminosities and Eddington ratios have revealed significant dispersion around the canonical scaling relation. In particular, the super-Eddington accreting massive black hole (SEAMBH) collaboration \citep{Du2014_SEAMBH, Du2015_SEAMBH, Du+2016, Du2018_SEAMBH, Hu2021_SEAMBH} and the Seoul National University (SNU) AGN monitoring project \citep{Woo+2023} have shown that high-accretion-rate AGNs systematically exhibit H$\beta$ lags shorter than those predicted by the canonical $R-L$ relation. Independent spectro-interferometric measurements have recently corroborated these findings, confirming that super-Eddington AGNs may indeed host more compact BLRs than previously assumed \citep{Gravity+2024, Gravity_Abuter+2024}. Concurrently, the Sloan Digital Sky Survey Reverberation Mapping (SDSS-RM) project \citep{Shen+2015, Grier+2017, Fonseca+2020, Shen+2024}, which targets non-local AGNs at $0.1<z<4.5$, has found a large intrinsic dispersion of $\sim 0.3$ dex around the mean H$\beta$ $R-L$ relation.
Recent efforts to refine the $R-L$ relation have sought to reduce this scatter by incorporating secondary parameters \citep{Du&Wang2019, Martinez-Aldama+2019, Fonseca+2020, Wang&Woo_2024}. For instance, the strength of the optical Fe II emission can tighten up the $R-L$ relation via the eigenvector 1 relations \citep{Boroson&Green1992, Sulentic+2000, Shen&Ho2014, Du&Wang2019, Yu+2020_RL}. Complementary to these empirical revisions, theoretical models have been proposed to explain these deviations through anisotropic continuum emission and geometric self-shadowing at high accretion rate \citep{Wang2014_selfshadow}.

In our previous work \citep[][hereafter \citetalias{WuQ+2025}]{WuQ+2025}, we investigated AGNs with low-to-moderate accretion rates ($10^{-3}\leq\LLEdd\leq0.5$), where the accretion flow is predominantly in the SSD regime. While we successfully reproduced the global $R-L$ relation with constant photoionization parameters, those models predicted longer lags for higher-Eddington-rate AGNs, in contrast to the observed trend. This discrepancy highlights the need for a more appropriate treatment of the ionizing radiation field and BLR physical conditions as the accretion approaches the Eddington limit. As the accretion rate approaches or exceeds the Eddington limit, the disk becomes advection-dominated \citep{Abramowicz1988}, and increases in scale height, and the puffed-up geometry of the inner accretion flow leads to self-shadowing that reduces the ionizing flux intercepted by the BLR clouds \citep{Wang2014_selfshadow}. A similar shielding scenario has been invoked to explain the weak high-ionization emission lines and X-ray weakness observed in weak-line quasars, in which a geometrically thick inner accretion disk blocks ionizing radiation from reaching the BLR and, along certain sightlines, the X-ray-emitting region \citep[e.g.][]{Wu_etal2011_WLQ, Wu_etal2012_WLQ, Luo_etal2015_WLQ, Ni+2018, Ni+2022}. Furthermore, the global gas supply and density within the central parsec increase with a higher accretion rate \citep{Hopkins+2024, Hopkins2025_FIRE}, potentially causing the line-emitting gas to migrate inward to maintain its ionization state.

In this work, we construct a self-consistent framework that incorporates slim disk structures, angle-dependent spectral energy distributions (SEDs), and BLR photoionization to investigate the extent to which these effects can explain the observed $R-L$ offsets for high-Eddington-accretion AGNs. The paper is organized as follows: in \S\ref{sec:method}, we describe our physically motivated slim-disk SED models and the photoionization calculations that account for the disk-BLR geometry. 
We discuss our results in \S\ref{sec:results} and present our conclusions in \S\ref{sec:conclusion}. Following the methodology of \citetalias{WuQ+2025}, we compare our theoretical results against an observational sample of $179$ broad-line AGNs with H$\beta$ RM measurements at $z\lesssim0.8$ \citep{Bentz2013, Bentz&Katz_2015, Du2015_SEAMBH, Grier+2017, Du&Wang2019, Hu2021_SEAMBH, Shen2024_SDSSRM}. 

\section{Methodology}\label{sec:method}

We construct a physically motivated model that combines AGN SEDs with photoionization calculations to study the impact of anisotropic illumination on BLR structure. Building on previous SED models (\citet{Kubota&Done2018, Kubota&Done2019}; hereafter \citetalias{Kubota&Done2018}, \citetalias{Kubota&Done2019}) and the locally optimally emitting clouds (LOC) framework \citep{Baldwin1977}, we incorporate angle-dependent ionizing continua from a slim disk geometry to compute two-dimensional line emissivities across a range of BH masses and accretion rates.
Throughout this work, accretion-disk calculations are performed in cylindrical coordinates. Disk quantities such as the scale height $H(R)$, effective temperature $T_{\rm eff}(R)$, and emitted continuum, are computed as functions of $R$.
The BLR clouds are instead described in spherical coordinates $(r,\theta)$, where $r$ is the distance from the central BH and $\theta$ is the viewing angle measured from the disk midplane unless otherwise noted.

\subsection{Incident AGN SEDs}

\subsubsection{Baseline SED models}

We base our SED models for AGN on a set of commonly-used packages {\tt agnsed}, {\tt qsosed}, and {\tt agnslim} (\citetalias{Kubota&Done2018}, \citetalias{Kubota&Done2019}), which compute the broadband continuum emission in terms of three principal components: (i) the outer, multi-temperature blackbody disk emission that dominates the optical–UV band; (ii) an intermediate, warm Comptonization component that contributes primarily to the soft X-ray excess; and (iii) the innermost, hot Comptonizing corona that produces the hard X-ray continuum. 
Model input parameters include the BH mass \Mbh, the dimensionless Eddington-scaled accretion rate $\log\dotm$, where $\dotm\equiv \dot{M}/\dot{M}_{\rm Edd} = \dot{M}/(L_{\rm Edd} \eta^{-1} c^{-2}$), source co-moving distance $D_{\rm c}$, inclination angle $\cos i$, the electron temperatures of the warm and hot Comptonizing regions $k T_e$, their spectral index $\Gamma$ and characteristic  radii $r_{\rm warm}$ and $r_{\rm hot}$. The original {\tt agnsed} models set the radiative efficiency to $\eta=0.057$ for a non-spinning BH, whereas in this work it is set to $\eta=0.1$. The viscosity is fixed at $\alpha=0.1$ throughout this work.

The {\tt agnsed} model represents the most basic configuration, adopting the Novikov–Thorne (NT) emissivity \citep{Novikov&Thorne1973} as the underlying accretion flow. The local disk emissivity at radius $R$ is expressed as:
\begin{equation}
    F_{\rm NT} (R) = \frac{3 G M_{\rm BH} \dot{M} f(R, a_\ast)}{8\pi R^3}=\frac{L}{4\pi R^2}\frac{3f(R, a_\ast)\Rg}{2\eta(a_\ast)R}
\end{equation}
where $\dot{M} [{\rm g\,s^{-1}}]$ is the mass accretion rate, $a_\ast$ is the dimensionless spin parameter, $L=\eta(a_\ast)\dot{M}c^2$ is the total integrated luminosity from the disk, $\eta(a_\ast)$ is the radiative efficiency, $R$ in units of $\Rg=GM_{\rm BH}/c^2$, and $f(R,a_\ast)$ is the standard NT correction factor accounting for the relativistic boundary condition at the innermost stable circular orbit. The hard X-ray emitting corona in this model is treated as a homogeneous spherical region within the truncated disk (\citetalias{Kubota&Done2018}; \citet{Hagen+2024}). Its total power is taken as the sum of the gravitational energy dissipated within the innermost accretion flow at $R<R_{\rm hot}$, and the luminosity of seed photons intercepted from the outer disk regions at $R>R_{\rm warm}$ that undergo inverse Compton scattering in the corona (see Equations 2–4 in \citetalias{Kubota&Done2018}).  

Compared to the original {\tt agnsed}, the simplified model {\tt qsosed} fixes the fraction ($2\%$) of energy dissipation into the hot corona $L_{\rm diss,\, hot} = 0.02L_{\rm Edd}$ and adopt typical values for the warm and hot Comptonization components as follows: $kT_{\rm e,\, hot} = 100\, \keV$, $kT_{\rm e,\, warm} = 0.2\, \keV$, $\Gamma_{\rm warm} = 2.5$, $R_{\rm warm} = 2R_{\rm hot}$, $R_{\rm out} = R_{\rm sg}$, $h_{\rm max} = 100$. 
In \citetalias{WuQ+2025}, we used this {\tt qsosed} model with fixed co-moving distance $D_c=1000\,{\rm Mpc}$, redshift $z=0$, and inclination $\cos{i}=0.5$ for the $\log\dotm\leq0$ AGN SEDs.
This model works generally well for AGNs with low-accretion rates and an intermediate SMBH mass $7.5<\log\Mbh<9.0$ compared to stacked observational data \citep{Mitchell_etal2023, Hagen+2024}.

At lower accretion rates ($\log\dot{m}\leq-1.5$), we do not include a warm Comptonization component. Observationally, the soft X-ray excess weakens substantially or disappears in low-luminosity AGNs, whereas the hard X-ray component often remains prominent \citep{Mitchell_etal2023, Hagen+2024}. This behavior is consistent with the disk-collapse picture discussed by \citet{Hagen+2024}, in which the inner UV-bright disk and warm-corona structure recede or collapse at low accretion rates, leaving a hotter, more radiatively inefficient inner flow dominated by the hard corona \citep{Yuan&Narayan2014}. We then approximate the low-$\dot{m}$ SED using only an outer disk and a hot corona, adopting $\Gamma_{\rm hot}=-1.7$ and $R_{\rm hot}=200\, R_{\rm g}$. In the standard {\tt qsosed} prescription, the dissipated hot-corona power is fixed to $L_{\rm diss,hot}=0.02L_{\rm Edd}$; however, at low $\dot{m}$, this can become comparable to or exceed the luminosity available from the accretion flow, and it also imposes an accretion-rate-independent hard-X-ray normalization. We therefore rescale the hard-corona dissipation as $L_{\rm diss,\, hot}/L_{\rm Edd} = 0.02 (\dotm/10^{-1.5})$ to preserve continuity with the standard {\tt qsosed} normalization at $\log\dotm=-1.5$ while allowing the hard-X-ray power to decline toward lower accretion rates.
 
At higher accretion rates, e.g., $\dotm=\dot{M}/\dot{M}_{\rm Edd}\gtrsim 0.3$, however, advection of thermal energy becomes dynamically important, and the flow begins to deviate from the SSD \citep{Abramowicz1988, Abramowicz_etal2013}. To account for this, \citetalias{Kubota&Done2019} introduced the {\tt agnslim} model as a simplified prescription for AGNs accreting near or above the Eddington limit. The principal modification lies in the treatment of the effective temperature profile: when the NT flux exceeds the local Eddington flux $F_{\rm Edd}(R) = L_{\rm Edd}/4 \pi R^2$, the model imposes a saturation, replacing $F_{\rm NT}(R)$ with $F_{\rm Edd}(R)$. In addition, \citetalias{Kubota&Done2019}  also allows the inner radius $R_{\rm in}$ to vary as a function of the accretion rate, producing a slab corona rather than a compact spherical corona at high accretion rates.  
Although this simplification has a limited impact on broadband SED fits due to parameter degeneracies, it intrinsically overestimates the effective disk temperature and cannot provide the disk height information, especially in the near-Eddington regime.
To address this, we implement an alternative approach in which the underlying temperature component is not parameterized through flux saturation, but instead computed by directly solving the governing slim disk equations that self-consistently incorporate advection, radial energy transport, and vertical structure effects.

\subsubsection{Slim disk SED model}

\begin{figure*}
    \includegraphics[width=\linewidth]{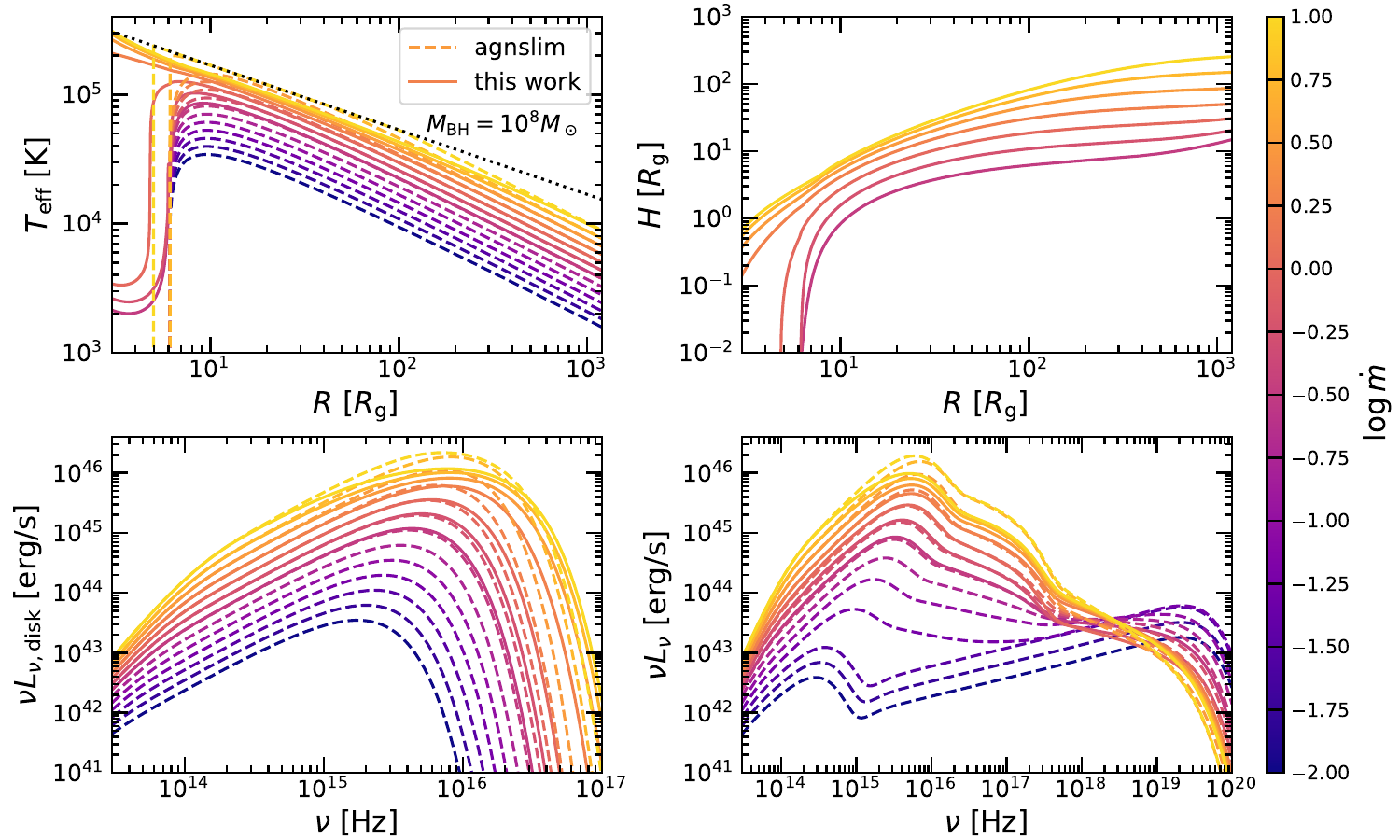}
    \caption{Comparison of the {\tt agnslim} (dashed) and the slim disks (solid) for a black hole of $\Mbh=10^8\,\msun$ with different accretion rates \dotm. 
    Top left panel: The effective temperature $T_{\rm eff}$ against radii $R$; the dotted black line shows the local Eddington flux $F_{\rm Edd}$. For the {\tt agnslim} model, the sharp vertical turnover for most solutions corresponds to the inner radius $R_{\rm in}$ where the disk terminates (see details for $R_{\rm in}$ in \citetalias{Kubota&Done2019}); while for our slim-disk solution, it indicates the radius at which the temperature exhibits a sharp decline.
    Top right: Scale height of the disk thickness.
    Lower left panel: the emergent spectra from the slim disk. 
    Lower right: the full SEDs with warm- and hot-corona components. Note when $\log\dotm\leq -1.5$ (the lowest three curves), the {\tt qsosed} model does not have the warm corona component, and the SED shape is completely different.
    }\label{fig:agnsed_vs_slim}
\end{figure*}

There have been a set of studies discussing the slim disk equations in detail \cite[e.g.][]{Abramowicz1988, Chen&Wang2004, Gu&Lu2007, Kato+2008, Jiao_etal2009, Wang2014_selfshadow, Feng+2019}. 
For simplicity, we use the vertically averaged slim-disk equations here. {We assume that the midplane density and temperature are related through a polytropic relation $p_0\propto\rho_0^{1+1/N}$ \citep{Hoshi1977}, and adopt $N=3$ throughout our calculations. The integration constants $I_N=(2^N N!)^2/(2N+1)!$ gives $I_{N=3}=16/35$ and $I_{N+1=4}=128/315$.} The surface density and vertically integrated pressure are then given by $\Sigma=2I_N\rho_0 H$ and $W=2I_{N+1}p_0H$, respectively. The disk half-thickness $H$ is determined from vertical hydrostatic equilibrium $\Omega_{\rm K}^2 H^2 = 2(N+1)p_0/\rho_0=(2N+3)W/\Sigma$, where $\Omega_K = (G\Mbh/R)^{1/2}/(R-2R_g)$ is the Keplerian angular momentum in a pseudo-Newtonian potential. The equation of state is $p_0 = p_{\rm rad} + p_{\rm gas}= 4\sigma_{\rm sb}T^4/3c + k_{\rm B} \rho_0 T/\mu m_{\rm p}$, where the first term on the right-hand side represents the radiation pressure ($\sigma_{\rm sb}$ is the Stefan–Boltzmann constant, $c$ is the speed of light, and $T$ is the mid-plane temperature), and the second term represents the gas pressure ($k_{\rm B}$ is the Boltzmann constant, $\mu = 0.617$ is the mean molecular weight, and $m_{\rm p}$ is the mass of the proton). 

The slim disk equations for the mass conservation, the radial component of the momentum conservation, and the angular momentum conservation are written as follows:
\begin{align}
    \dot{M} &=-2\pi R \Sigma v_R, \\
    v_R\frac{dv_R}{dR}+\frac{1}{\Sigma}\frac{d W}{dR} &=\frac{\ell^2-\ell_{\rm K}^2}{R^3}-\frac{W}{\Sigma}\frac{d\ln \Omega_{\rm K}}{dR}, \\
    \dot{M}(\ell-\ell_{\rm in}) &=-2\pi R^2 T_{r\phi}, 
\end{align}
where $v_R$ is the radial velocity, $\ell\equiv Rv_\psi$ and $\ell_{\rm K}\equiv R^2\Omega_{\rm K}$ is the specific angular momentum and the Keplerian angular momentum of the gas, and $T_{r\phi}=-\alpha W$ is the turbulent viscosity. 

The energy equations of the slim disk are:
\begin{equation}
Q_{\rm vis}=Q_{\rm adv} + Q_{\rm rad},
\end{equation}
which includes the gravitational power released by viscosity $Q_{\rm vis}$, the advective cooling $Q_{\rm adv}$, and radiative cooling using the diffusion approximation $Q_{\rm rad}$. The explicit form of the heating and cooling rates is given by:
\begin{equation}
Q_{\rm vis}=RT_{r\phi}\frac{d\Omega}{dR}.
\end{equation}
\begin{gather}
Q_{\rm adv}=\frac{\dot M}{2\pi R}\frac{W}{\Sigma}\xi,\\
\begin{split}
\xi = &-\frac{\Gamma_1+1}{2(\Gamma_3-1)}\frac{d\ln{W}}{d R} + \frac{3\Gamma_1-1}{2(\Gamma_3-1)}\frac{d\ln{\Sigma}}{d R} \\
&+ \frac{\Gamma_1-1}{\Gamma_3-1}\frac{d\ln{\Omega_K}}{d R}
\end{split}
\end{gather}
\begin{align}
    \Gamma_1 & = \beta + \frac{(\gamma-1)(4-3\beta)^2}{\beta+12(\gamma-1)(1-\beta)} \\
    \Gamma_3 - 1 & = \frac{(\gamma-1)(4-3\beta)}{\beta+12(\gamma-1)(1-\beta)} = \frac{\Gamma_1-\beta}{4-3\beta}  
\end{align}
\begin{equation}
Q_{\rm rad}=\frac{32\sigma_{\rm sb}T^4}{3\tau}, 
\end{equation}
where $\Gamma_1$ and $\Gamma_3$ are functions of the ratios of the specific heats and gas pressure fraction $\beta$, and $\tau=(\kappa_{\rm es}+\bar{\kappa}_{_{\rm ff}})\Sigma$ is the optical depth, here $\kappa_{\rm es}=0.40$ is the opacity of electron scattering and the Rossland mean opacity $\bar{\kappa}_{_{\rm ff}}=6.4\times 10^{22}\rho_0 T^{-3.5}$ \citep{Kato+2008}.

The above equations governing the slim disk can be reduced to a system of two coupled, first-order differential equations. We integrate these equations numerically with the fourth-order Runge-Kutta method, starting from an outer boundary at $10^{4}\Rg$, where the disk structure is assumed to approach the SSD solution. The integration is carried inward to an inner boundary at $R_{\rm in} = 3\Rg$, with the specific angular momentum at the inner edge $\ell_{\rm in}$ determined from the shooting method. Its value is adjusted iteratively with the shooting method until a global, transonic solution is obtained \citep{Chen&Wang2004}.

With the global structure of the slim disk solved, we calculate the effective temperature $T_{\rm eff}$ as a function of the radius:
\begin{equation}\label{eq:Teff}
    T_{\rm eff} (R) = 
    \left(\frac{Q_{\rm rad}}{2 \sigma_{\rm sb}}\right)^{1/4}
\end{equation}
The local emissivity profile in {\tt qsosed} is substituted by the effective temperature with $F(R)=\sigma_{\rm sb} T_{\rm eff}^4$. 
The SED generation code used in this work is publicly available online \footnote{\url{https://github.com/QiaoyaWu/agnsed_python}}.

Figure~\ref {fig:agnsed_vs_slim} summarizes the solution of the slim disk and compares the different effective temperatures and spectra between the agnslim model and our slim disk implementation. 
In the top left panel, the default {\tt agnslim} temperatures are systematically higher than the slim-disk solution, leading to an overestimated $F(R)$. This behavior is consistent with advection and photon trapping in slim disks, which reduce the radiative surface temperature at high \dotm.
The top right panel of the scale height profile indicates that as \dotm\ increases, radiation pressure and advection inflate the inner flow, increasing the scale height $H$.
The lower panels demonstrate the emergent spectra of the original \citetalias{Kubota&Done2019} model with those computed using the revised temperature profiles. In the original model, increasing \dotm\ raises the overall SED, shifts the UV peak to higher frequencies, and weakens the hard-X-ray component. When the slim-disk temperatures are adopted, the cooler disk temperature suppresses the UV bump and attenuates the blueward shift, yielding a relatively softer optical–UV continuum at the same \dotm. For the lowest accretion-rate models ($\log\dotm\leq-1.5$), the SED shape differs from the higher-$\dot{m}$ cases because we remove the warm Comptonization component and approximate the inner flow only as a hot corona.

We note that for $\log\dotm=-0.5$, the temperature profiles between these two models agree beyond $R\gtrsim10\Rg$, and inner-radius differences have a negligible impact on the warm and disk components. Thus, using this improved AGN SED framework, we compute a grid of SEDs for non-spinning black holes ($a_\ast=0$) spanning $M_{\rm BH}=10^{7}$–$10^{9}\,\msun$ and dimensionless accretion rates $-2\le\log\dot m\le 1$. 
We adopt {\tt qsosed} for sub-Eddington cases ($\log\dotm<-0.5$) and solve the slim-disk equations for near- and super-Eddington cases ($\log\dotm\ge-0.5$). At the transition point ($\log\dotm=-0.5$), the two models produce nearly identical SEDs, ensuring a smooth and continuous connection across the grid. The grid contains $5\times13$ logarithmically spaced SEDs with step sizes of $0.5$ dex in $\log M_{\rm BH}$ and $0.25$ dex in $\log\dot m$. 

\begin{figure*}
    \includegraphics[width=\linewidth]{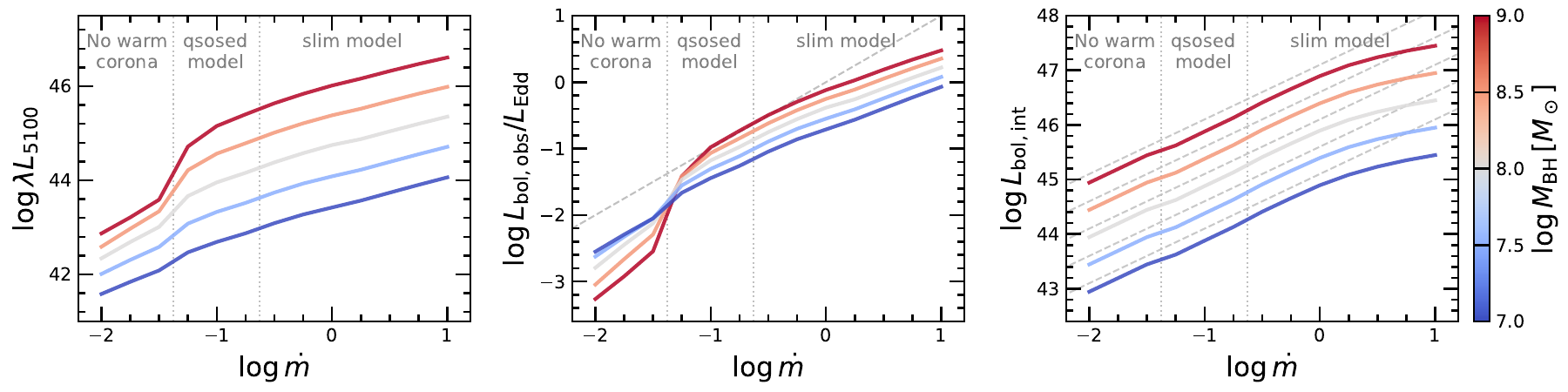}
    \caption{Dependence of the observed optical luminosity $\lambda L_{5100}$ and Eddington ratio on \dotm. The vertical dotted lines mark transitions between different accretion regimes: low-accretion {\tt qsosed} models without warm corona, standard {\tt qsosed} models, and slim-disk models at high \dotm. Curves are color-coded by BH mass, as indicated by the color bar. 
    Left: Monochromatic luminosity $\Lopt$ computed from the model SEDs as a function of \dotm, assuming the observer receives the full SED without shadowing. 
    Middle: Corresponding Eddington ratio estimated from optical luminosity $\LobsLEdd\equiv(\Lopt\times9.26)/(1.26\times10^{38}\Mbh)$ as a function of \dotm. 
    Right: Bolometric luminosity integrated from the model SEDs $L_{\rm bol, int}=\int L_\nu\,d\nu$. The dashed gray lines indicate the corresponding linear scaling with $\dotm L_{\rm Edd}$.
    }\label{fig:mdot_mapped_obs}
\end{figure*}

Because direct estimates of the dimensionless accretion rate \dotm\ are uncertain for individual AGNs, we use the Eddington ratio based on the optical continuum luminosity and a constant bolometric correction factor $\LobsLEdd\equiv(\Lopt\times 9.26)/(1.26\times10^{38} \Mbh)$ as a proxy for both observational and theoretical samples to ensure a fair comparison \citep{Richards+2006}. Figure~\ref{fig:mdot_mapped_obs} illustrates the mapping between the intrinsic accretion rate $\log \dotm$ in our models and the corresponding observable quantities.
The optical and bolometric luminosities increase approximately linearly with \dotm, while as the accretion rate approaches the Eddington limit, the relations become progressively flatter and deviate from a simple linear scaling. This occurs because photon trapping reduces the radiative efficiency, causing an increasing fraction of the dissipated energy to be advected inward rather than escaping as radiation \citep{Abramowicz1988, Watarai+2000, Sadowski+2014}. As a result, the observed luminosity grows more slowly than the intrinsic mass accretion rate, producing a luminosity saturation effect at high $\dotm$. The mild curvature near the transition between the no-warm-corona and {\tt qsosed} regimes likely reflects the sensitivity of the modeled optical-to-EUV SED shape to the adopted SED prescription. This effect is most pronounced for high-\Mbh\ systems, whose cooler disks shift the thermal SED peak toward longer wavelengths \citep{Temple+2023}, making $\Lopt$ more sensitive to changes in the disk and warm-corona components. 

\subsection{Photoionization with self-shadowing}\label{sec:self_shadow_compute}

\subsubsection{Shielding geometry}
\begin{figure}
    \includegraphics[width=\linewidth]{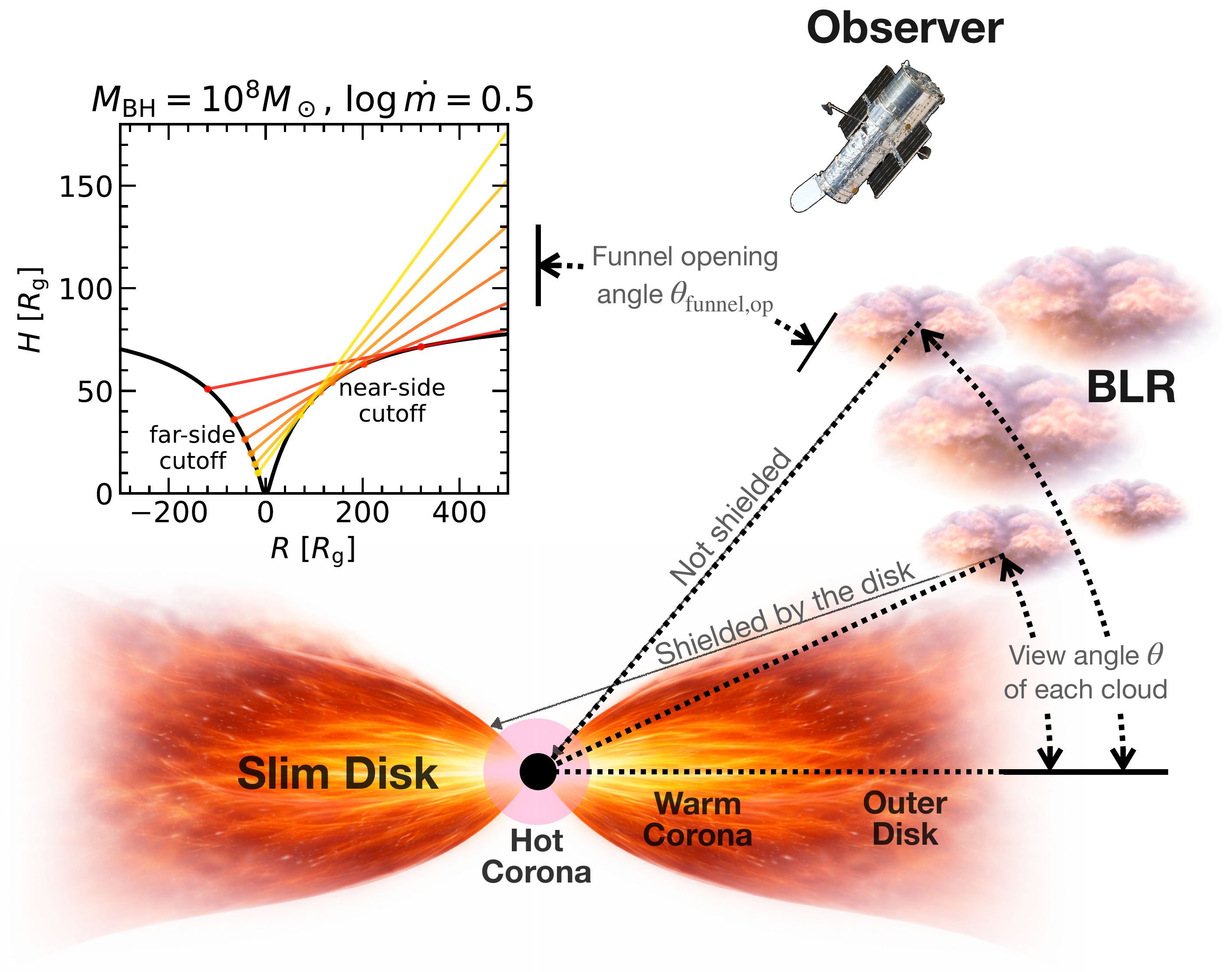}
    \caption{Schematic of self‐shadowing in a slim-disk AGN. The puffed-up inner disk forms a funnel, producing an anisotropic radiation field. The funnel opening angle $\theta_{\rm funnel, op}$ defines the polar region devoid of BLR clouds and is measured from the polar axis, while the viewing angle $\theta$ denotes the angle between the cloud position and the disk midplane.
    BLR clouds at different viewing angles therefore receive different ionizing fluxes and continuum shapes. 
    For a cloud at a given $(r,\theta)$ location, the tangent line from the cloud to the disk surface defines the near-side cutoff radius, interior to which the near side of the disk is hidden from the cloud; extending this tangent line to the opposite side of the disk defines the far-side cutoff radius, within which the far side of the disk is not visible. These tangent lines are used only to determine the portions of the disk contributing to the incident continuum and to illustrate the self-shadowing geometry.
    The inset panel in the upper left shows the near- and far-side cutoff radii produced by self-shadowing for BLR clouds located at $r=10^4\,\Rg$, and the colored lines correspond to clouds at viewing angles $\theta=3^\circ$, $6^\circ$, $9^\circ$, $12^\circ$, $15^\circ$, and $18^\circ$ (from red to yellow).
 }\label{fig:scheme}
\end{figure}

\begin{figure}
    \includegraphics[width=\linewidth]{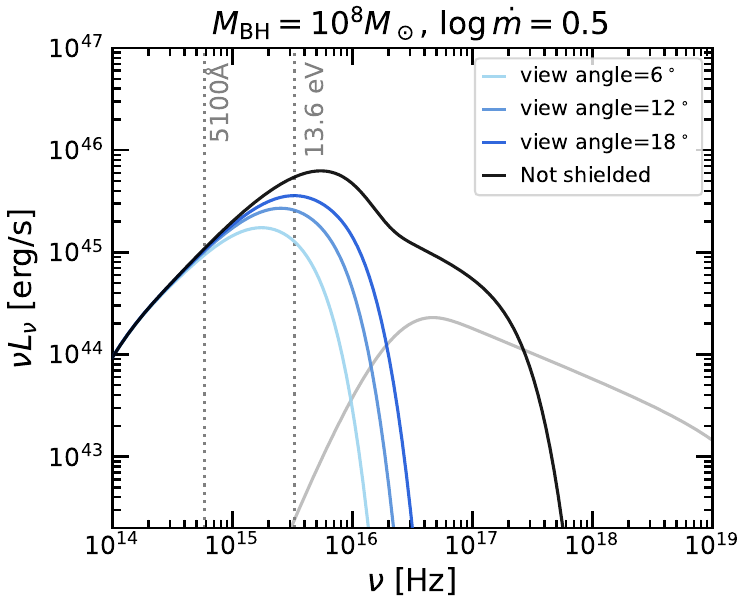}
    \caption{Angle-dependent AGN SED (disk+warm components) for different viewing angles at $r=10^4\Rg$, illustrating the effects of accretion-disk self-shadowing. The plotted SEDs are the averages of the intercepted continua from the near and far sides. The grey line demonstrates the hot X-ray component. }\label{fig:sed_self_shadow}
\end{figure}

As discussed in \citet{Wang2014_selfshadow}, the inner region of the disk would form a funnel-like region, which produces an anisotropic radiation field. Because BLR clouds located at different radii ($r$) and viewing angles ($\theta$) are exposed to intrinsically different incident SEDs, to model the self-shadowing effect, we adopt a local tangent line $dH/dR$ to determine the near-side cut-off radius on the disk surface for a cloud at any given position. This line is then extended to define the far-side cutoff region, as illustrated in the schematic in Figure~\ref{fig:scheme}. Given that the geometry of the optically thin hot corona remains uncertain, we do not explicitly model its shadowing and instead retain the hard X-ray components. 
Although the slim disk funnel is a three-dimensional structure and both the tangency and cut-off radii vary with azimuth angle $\phi$, we approximate the geometry with an azimuthally averaged meridional cross-section, i.e., for each cloud position, we adopt the average of the near- and far-side intercepted SEDs. We do not additionally include disk inclination projection effects, since the angular dependence of the emitted radiation from a geometrically thick slim disk remains uncertain. Moreover, in the self-shadowed regions, the reduction in ionizing flux due to geometric obscuration is expected to dominate over the gradual projection effect.
Figure~\ref{fig:sed_self_shadow} illustrates the resulting anisotropy of the emergent SED, arising from the slim-disk self-shadowing. Self-shadowing preferentially attenuates low-latitude (near-equatorial) sightlines, reducing the EUV/soft X-ray flux from the warm corona and inner disk regions.

\begin{figure*}
    \includegraphics[width=\linewidth]{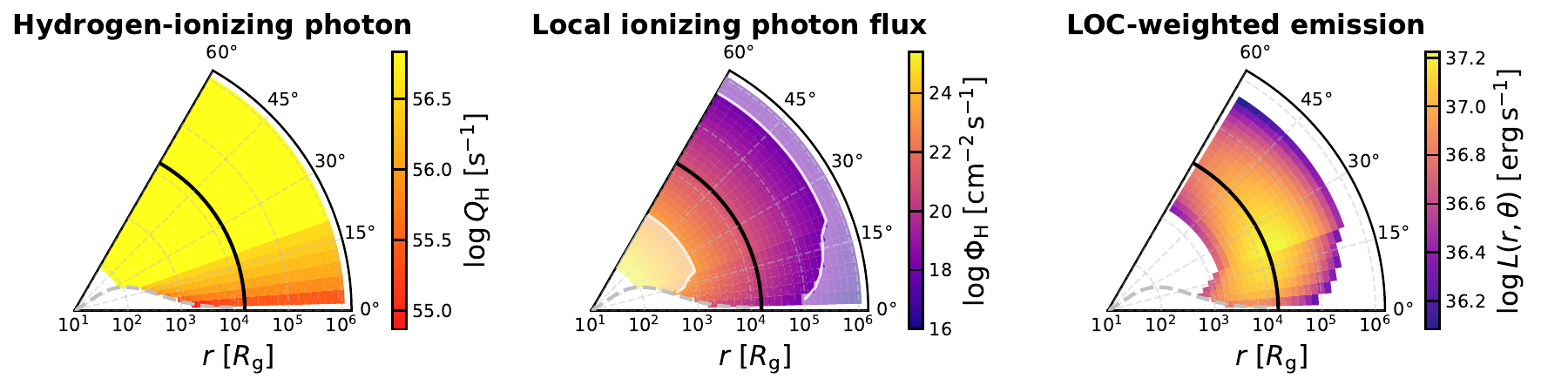}
    \caption{Polar maps of the BLR photoionization illustrating the angular and radial dependence of the incident radiation field for a $\Mbh=10^8\msun$ and $\log\dotm=0.5$ AGN. Black curves indicate the emissivity-weighted BLR radius, and the gray dashed curves show the vertical structure of the accretion-disk surface.
    Left: Hydrogen-ionizing photon rate $Q_{\rm H}$ received by clouds. 
    Middle: Local ionizing photon flux $\Phi_{\rm H}$ at each position. The whitened innermost and outermost illuminated regions are excluded from the LOC integration: outer regions where dust grains can survive ($\PhiH<10^{18}\,{\rm cm^{-2}\,s^{-1}}$), and all regions where $\Phi$ falls outside the allowed range corresponding to $6.25 \leq \log(Uc) \leq 11.25$.
    Right: LOC-weighted emission $L(r, \theta)$ after integrating over gas density and radial gas weighting.}
    \label{fig:BLR_photoionization_grid}
\end{figure*}

\begin{figure*}
    \includegraphics[width=\linewidth]{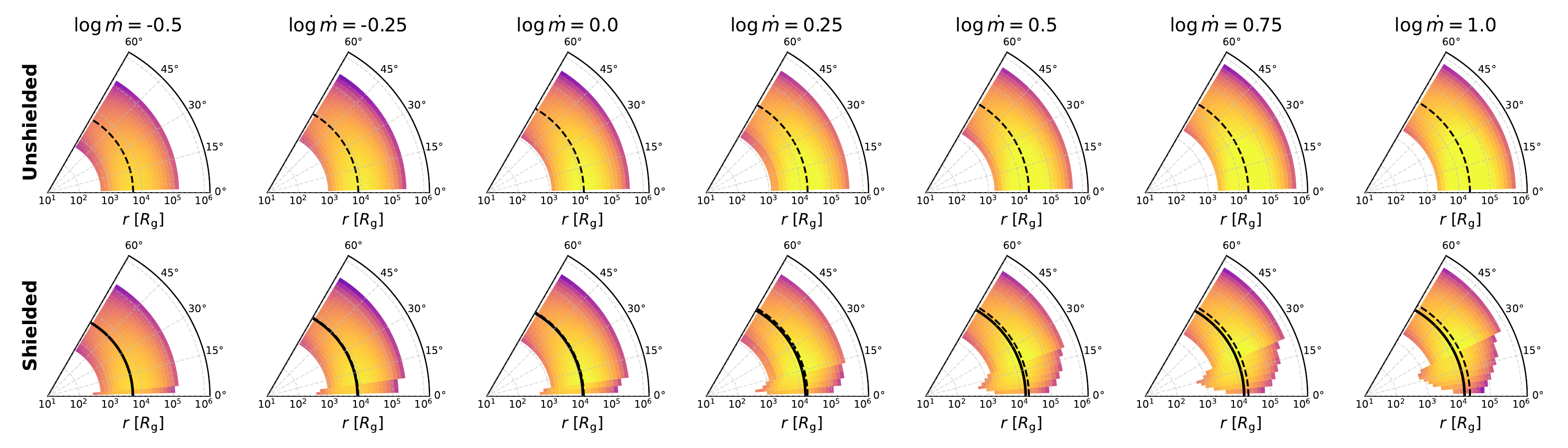}
    \caption{Weighted BLR emission distribution in the $(r,\theta)$ plane for $10^8\,\msun$ BHs with different accretion rates ($\log\dotm = -0.5$ to $1.0$). The dashed curves mark the emissivity-weighted BLR radius without self-shadowing effects, while the solid curves indicate the corresponding characteristic radius after applying self-shadowing.}
    \label{fig:loc_shield_vs_unshield}
\end{figure*}

Consequently, we assume an axisymmetric BLR distribution above the disk and construct a two-dimensional grid in radius and polar angle (Figure~\ref{fig:BLR_photoionization_grid}) for photoionization calculations.
Observational and BLR dynamical modeling studies generally indicate that the BLR is not concentrated near the polar axis \citep[e.g.,][]{Wills1986, Shen&Ho2014, Mejia-Restrepo2018, Pancoast_2014_lick}; instead, it occupies intermediate latitudes, with a typical opening angle $\theta_{\rm BLR} \sim 30^\circ-60^\circ$ from the disk midplane \citep{Williams+2018, Gravity+2024, Gravity_Abuter+2024, ZStone+2025}. In addition, the physical extent of the BLR is expected to be of a similar order of magnitude to the cloud distances inferred from RM measurements and to remain within the dust sublimation radius \citep{Netzer&Laor1993, Goad+2012}. Guided by these considerations, we construct a BLR spatial grid $(r, \theta)$ over radii $10^2$ to $10^6\,\Rg$ and viewing angles $\theta \in[ 0^\circ, 60^\circ]$ measured from the disk plane (BLR opening angle $\theta_{\rm BLR}=60^\circ$ and funnel opening angle $\theta_{\rm funnel, op}=30^\circ$). BLR clouds located inside the accretion disk are not included in our computation. In our slim-disk model, the disk structure is computed out to an outer radius of $R_{\rm out} = 10^4\Rg$: for $r\leq R_{\rm out}$, we exclude cells that lie within the disk body, defined by $r\sin{\theta} \leq H(R)$, where $H(R)$ is the local disk scale height; for $r > R_{\rm out}$, we adopt a constant disk thickness at the outer boundary and exclude cells satisfying $r\sin\theta < H(R_{\rm out})$.

\subsubsection{CLOUDY and LOC model}

We compute BLR line emissivities with {\tt CLOUDY} (\cite{Cloudy25}, version 25.0) under the LOC framework \cite[e.g.][]{Baldwin+1995, Korista+1997, Ferguson+1997, Korista&Goad_2000, Korista&Goad_2004, Korista&Goad2019, Guo+2020}. 
Following \citetalias{WuQ+2025}, each model illuminates plane-parallel clouds of solar abundance ($Z=Z_\odot$) and hydrogen column density $N_{\rm H}=10^{23}\,\rm{cm^{-2}}$ in an open geometry. The cloud photoionization computations span hydrogen number density $8\le\log \nH\,[{\rm cm^{-3}}]\le16$ in $0.5$ logarithmic steps and incident hydrogen-ionizing photon flux $16\le\log\PhiH\,[{\rm cm^{-2}\,s^{-1}}]\le25$ in logarithmic steps of $0.25$ dex for each SED.

We first compute the hydrogen-ionizing photon rate \QH\ via the angle-dependent SED incident on the cloud at each $(r,\theta)$ cell:
\begin{equation}
     \QH = \int^\infty_{\nu_{\rm H}}\frac{\pi L_{\nu}}{h\nu}\,d\nu
\end{equation}
from which the incident ionizing photon flux follows as $\PhiH={\QH}/{4\pi R^2}$. We show the two-dimensional distributions of both \QH\ (left) and \PhiH\ (middle) in Figure~\ref{fig:BLR_photoionization_grid}. The photoionization parameter $U$ is then related to the hydrogen number density $n_{\rm H}$ via $U = \PhiH/\nH c$ in each cell. 

\begin{table}
\centering
\begin{tabular}{ccccccc}
\hline\hline
\multicolumn{7}{c}{Median $\log L_{\rm H\beta}$ [$\ergs$]} \\ \hline
$\Gamma$ & \multicolumn{6}{c}{BLR opening angle $\theta_{\rm BLR}$} \\
\cline{2-7} & $15^\circ$ & $30^\circ$ & $45^\circ$ & $\mathbf{60^\circ}$ & $75^\circ$ & $90^\circ$ \\
\hline
$-1.6$ & 41.24 & 41.77 & 41.94 & 42.01 & 42.05 & 42.07 \\
$\mathbf{-1.5}$ & 41.73 & \textbf{42.22} & \textbf{42.40} & \boxed{\textbf{42.48}} & \textbf{42.52} & \textbf{42.55} \\
$-1.4$ & 42.19 & 42.67 & 42.81 & 42.89 & 42.93 & 42.95 \\
$-1.3$ & \textbf{42.61} & 43.12 & 43.26 & 43.34 & 43.38 & 43.40 \\
$-1.2$ & 43.10 & 43.57 & 43.72 & 43.79 & 43.83 & 43.85 \\
$-1.1$ & 43.52 & 44.03 & 44.17 & 44.25 & 44.29 & 44.31 \\
$-1.0$ & 44.02 & 44.46 & 44.62 & 44.73 & 44.78 & 44.81 \\
\hline
\hline
\end{tabular}
\caption{Median H$\beta$ line luminosity ($\log L_{\rm H\beta}$) predicted by the LOC model. The boxed value indicates the fiducial constant-density model adopted in this work, which best matches the observed median \hbeta\ luminosity ($\log L_{\rm H\beta}=42.41$). Boldface entries mark the models that provide the best match to the observed \hbeta\ luminosity \citep{Wu&Shen2022} at each BLR opening angle.}
\label{tab:median_lhb_cf_gamma}
\end{table}

\begin{figure}
    \includegraphics[width=\linewidth]{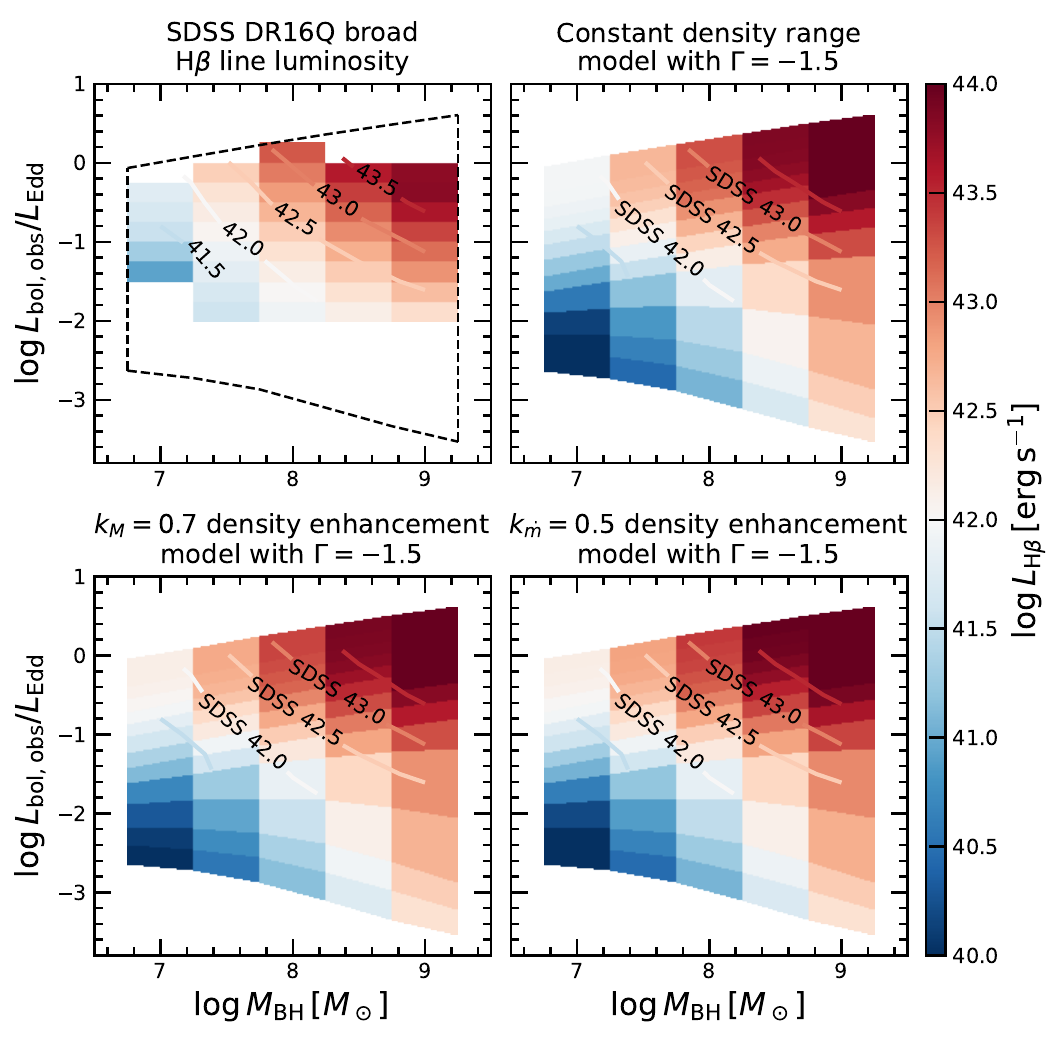}
    \caption{Comparison of the H$\beta$ line luminosity distribution in the $\Mbh-\LLEdd$ plane. The top-left panel shows the median SDSS DR16Q broad H$\beta$ line luminosity at each BH parameter \citep{Wu&Shen2022}, while the other panels show LOC model predictions for a fiducial constant-density-range BLR (top right), a BH-mass-dependent density enhancement with $k_M=0.7$ (bottom left), and an accretion-rate-dependent density enhancement with $k_{\dot{m}}=0.5$ (bottom right), all assuming $\Gamma=-1.5$. Colors indicate $\log L_{\rm H\beta}$ in units of \ergs. Contours show the SDSS median H$\beta$ luminosity levels overlaid on each panel for comparison. The dashed boundary in the SDSS panel indicates the coverage of our LOC model grid.}\label{fig:Lline_hbeta_gamma}
\end{figure}

The spatial distribution of the line-emitting gas is not known a priori. In the standard LOC framework, the emitting clouds are typically assumed to follow a spherically symmetric distribution. The total line luminosity is expressed as
\begin{equation}
L_{\rm line} \propto CF \iint r^2 \epsilon(\Phi,n) f(r) g(n)\,dn\,dr
\end{equation}
where $CF$ is the global covering factor of the BLR, and $\epsilon(\Phi, n)$ is the emission intensity from a single cloud surface, which depends on the incident ionizing flux $\Phi$ and gas density $n$. For an isotropic continuum source, $\Phi$ depends only on radius and therefore often serves as a proxy for distance. The function $f(r)$ describes the radial covering distribution of clouds, commonly parameterized as $f(r)\propto r^\Gamma$ with $\Gamma\in[-1.6, 1.0]$ depending on $CF$, while the gas density distribution is taken to be a power law $g(n)\propto n^\beta$ with $\beta=-1$, corresponding to equal weighting per logarithmic density interval \citep[e.g.][]{Baldwin+1995, Ferguson+1997, Korista&Goad_2000}.
However, the ionizing continuum in our model is anisotropic, so the emissivity must be evaluated on two-dimensional $(r,\theta)$ cells rather than in a spherically symmetric geometry. 
We retain $CF=50\%$ as a global normalization of the total line emission, while neglecting explicit cloud–cloud shadowing to isolate the effect of anisotropic disk illumination on the BLR emissivity distribution.
We also adopt the same LOC prescription for the density distribution $g(n)\propto n^{-1}$ \citep{Baldwin+1995, Ferguson+1997} and compute the emission contribution from each $(r, \theta)$ cell as:
\begin{equation}\label{eq:loc_integral}
\begin{split}
    L(r, \theta)  & \propto CF\, r^2 \cos{\theta}\,f(r)\int\epsilon(\Phi, n)g(n) dn \\
                  & \propto  CF\, r^{2+\Gamma} \cos{\theta}\int\epsilon(\Phi, n) d(\log{n})\\
                  &\propto  CF\, r^{2+\Gamma} \cos{\theta}\;\bar{\epsilon}(\Phi)
\end{split}
\end{equation} 
where $\bar{\epsilon}(\Phi)$ denotes the density-integrated emissivity, evaluated by interpolating over the computed photoionization density range. 
Our fiducial calculations integrate over the typical BLR density range $8\leq \log \nH \leq 12$ and require $6.25\leq \log Uc\leq11.25$ to restrict the integration to gas with ionization conditions appropriate for efficient BLR line emission, excluding highly ionized gas and nearly neutral regions. We also exclude cells with $\PhiH<10^{18}\,{\rm cm^{-2}\,s^{-1}}$, where dust survival is expected to suppress the line emission \citep{Korista&Goad_2000, Korista&Goad_2004}.
Following the LOC prescription of \citet{Korista&Goad_2000}, we explore different values of the radial index $\Gamma$ in the radial covering distribution and compute the resulting \hbeta\ luminosity. 
The total \hbeta\ line luminosity is estimated as
$L_{\rm H\beta} \propto \iint L(r,\theta)\,dr\,d\theta$. 
The results are summarized in Table~\ref{tab:median_lhb_cf_gamma}. We compare the model predictions with quasars from \citet{Wu&Shen2022} matched in BH mass and accretion rate, for which the median observed luminosity is $\log(L_{\rm H\beta}/{\ergs})=42.41$. Based on this comparison, we adopt $\Gamma=-1.5$ for our fiducial model with a BLR opening angle of $\theta_{\rm BLR}=60^\circ$.
The two-dimensional distribution of $L_{\rm H\beta}$ in the $\Mbh-\LLEdd$ plane is shown in Figure~\ref{fig:Lline_hbeta_gamma}. 
The resulting fiducial constant-density-range model (top right panel) captures the overall increase of $L_{\rm H\beta}$ toward higher $\Mbh$ and $\LLEdd$, and reproduces both the observed normalization and the gradients across the parameter space. We also consider different BLR opening angles and alternative density ranges to allow for systematically higher BLR gas densities in \S\ref{sec:R-L}.

We then determine the average BLR distance from the central SMBH based on the two-dimensional emissivity distribution. The computation procedure is illustrated in Figure~\ref{fig:BLR_photoionization_grid}. Additionally, Figure~\ref{fig:loc_shield_vs_unshield} presents the spatial distribution of the weighted BLR emission in the $(r,\theta)$ plane for $10^8\msun$ BHs in the near-Eddington accretion regime. In the unshielded case (top row), the emissivity-weighted BLR radius (dashed curves) varies only modestly with accretion rate, indicating that in the absence of anisotropy, the radial structure of the BLR is primarily governed by the radial dependence of the ionizing flux.
When self-shadowing is included (bottom row), the emission distribution becomes increasingly anisotropic with increasing accretion rate. The puffed-up inner disk blocks ionizing photons from reaching low-latitude regions, suppressing emission from clouds at small $\theta$. As a result, the effective line-emitting region shifts inward, leading to a systematic decrease in the emissivity-weighted BLR radius relative to the unshielded case. This effect becomes more pronounced at higher \dotm, where the disk thickness and degree of shielding are enhanced.

\section{Results and Discussion}\label{sec:results}
\subsection{The offset in the $R-L$ relation}\label{sec:R-L}
\begin{figure}
    \includegraphics[width=\linewidth]{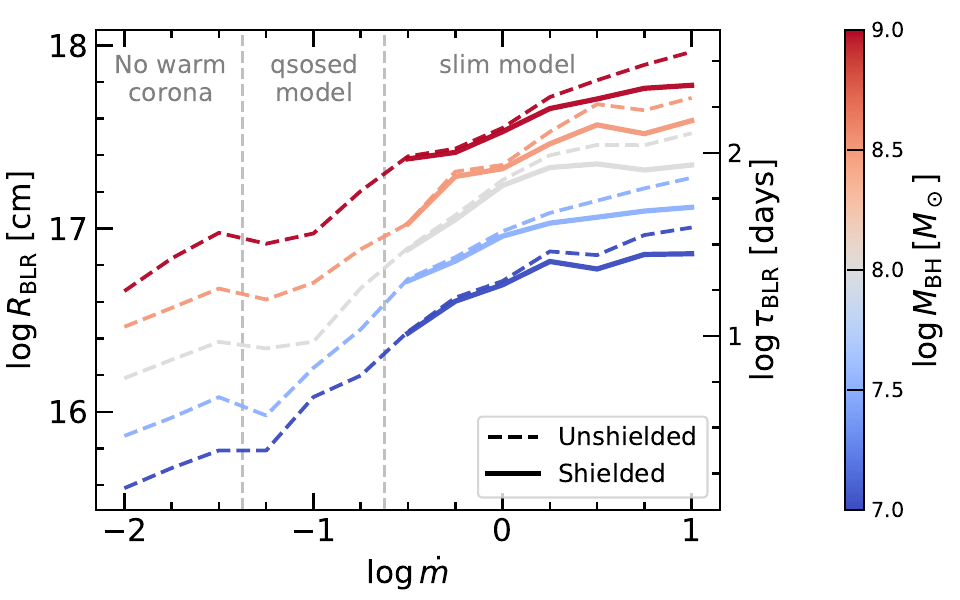}
    \caption{The inferred average BLR radius as a function of accretion rate. The dashed lines represent results from the {\tt agnslim} model without self-shadowing, while the solid lines correspond to the slim disk model including self-shadowing effects. Curves and data points are color-coded by SMBH mass. 
    }\label{fig:Rblr_mdot}
\end{figure}

Our previous work \citepalias{WuQ+2025} successfully reproduced the global $R-L$ relation in the sub-Eddington regime using universal values of hydrogen density ($\nH= 10^{12}$ cm$^{-3}$) and ionization parameter ($\log U = -2$); however, it predicted longer lags for higher Eddington rate AGNs, contradicting the observed trend. This discrepancy implies that the SED model we adopted does not correctly resolve the soft-UV part, and the assumption of constant gas density may not be realistic. 
In this work, we treat the anisotropic ionizing radiation field and BLR physical conditions consistently as the system approaches the Eddington limit in accretion. We solve the slim-disk structure equations and construct a location-dependent photoionization model to estimate the characteristic distance between the ionizing source and the BLR clouds, as described in \S\ref{sec:self_shadow_compute}. 
Figure~\ref{fig:Rblr_mdot} presents the predicted BLR radius as a function of accretion rate. The unshielded models yield systematically larger BLR radii than the self-shadowed models, whereas geometric shielding reduces the ionizing flux reaching the line-emitting gas at low viewing angles. This difference increases toward higher \dotm, where the inflated inner disk enhances self-shadowing and suppresses the effective BLR radius by up to $\sim0.2$ dex. Such a reduction in BLR size at high accretion rates is broadly consistent with the expectation from slim-disk self-shadowing \citep{Wang2014_selfshadow}.

\begin{figure*}
    \includegraphics[width=\linewidth]{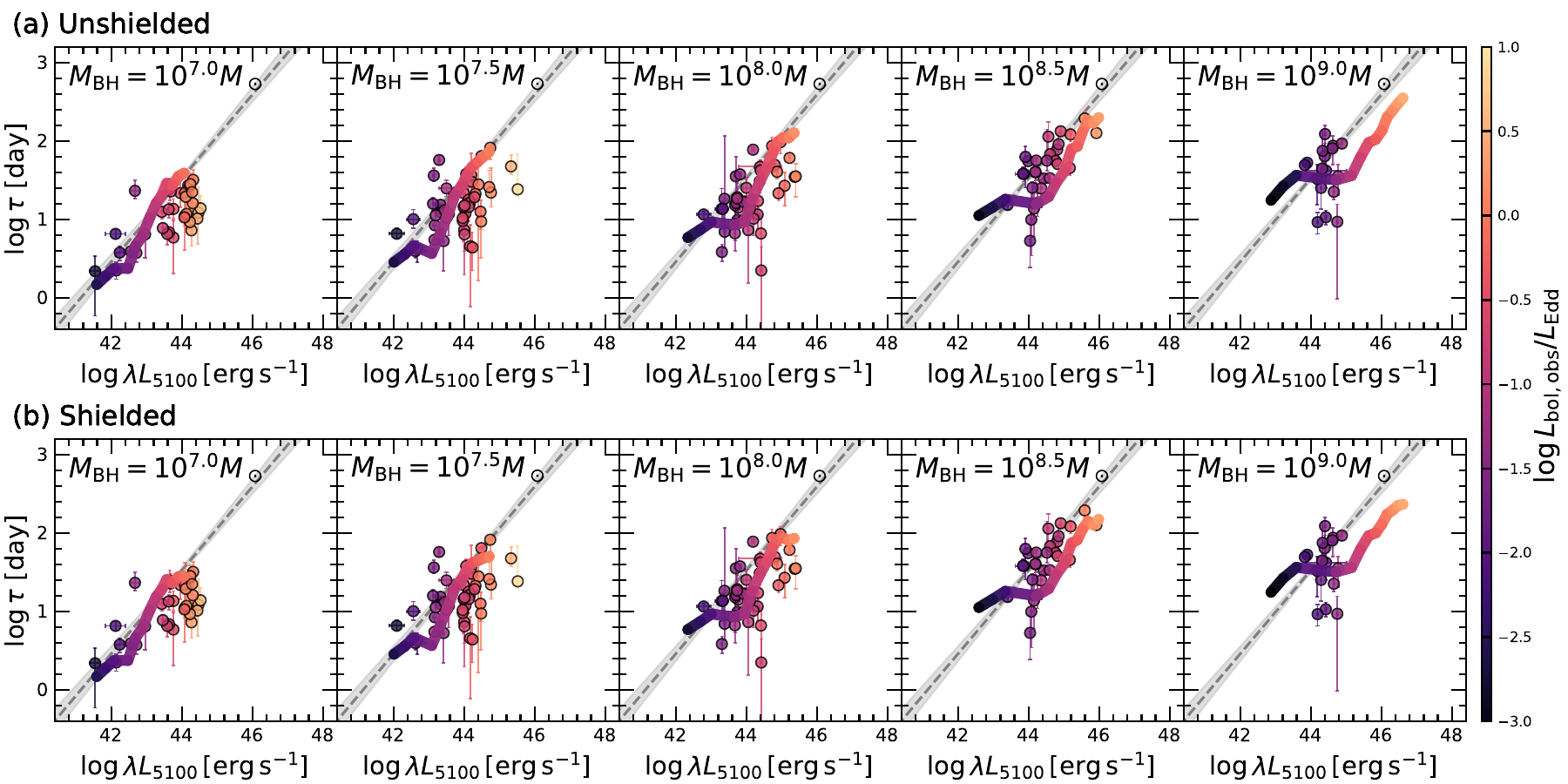}
    \caption{Dependence of the \hbeta\ BLR size on monochromatic luminosity at 5100 \AA\ across a range of mass bins (from $\log\Mbh=7$ to $\log\Mbh=9$ with $\Delta\log\Mbh=0.5$). The model predictions show the \hbeta\ emissivity-weighted BLR radius, expressed in light-days and used as an approximate proxy for the RM time lag for the unshielded case (top (a)) and shielded case (bottom (b)). Each curve corresponds to a fixed BH mass, while the color variation along the curve denotes the observed Eddington ratio \LobsLEdd. Observed \hbeta\ time lags are presented as circles with the same color scheme as the theoretical predictions. The gray dashed line indicates the canonical $R-L$ relation from \cite{Bentz2013}, with a slope $k = 0.533$ and an intercept $b=1.527$ at $\log(\Lopt/\ergs)=44$.
    }\label{fig:R-L}
\end{figure*}

Figure~\ref{fig:R-L} shows the predicted \hbeta\ BLR time lag versus the monochromatic luminosity at $5100$ \AA\ for different BH masses with $\Delta\log\Mbh=0.5$. 
Within each \Mbh\ bin, the predicted lag increases with $\LLEdd$ at low accretion rates, broadly following the observed normalization and slope of the empirical $R-L$ relation. As the accretion rate approaches the Eddington limit, however, changes in the disk temperature profile become less pronounced in the slim-disk regime, while geometric thickening of the inner disk suppresses the increase in the effective ionizing flux incident on BLR clouds. As a result, the growth of the effective BLR radius slows down even as the optical luminosity continues to increase, producing a flatter $R-L$ trend at high-\LLEdd.
Comparing the unshielded and shielded cases shows that self-shadowing primarily affects the high-accretion-rate end of the model sequences. The shielded models predict somewhat smaller \hbeta\ emissivity-weighted radii at high $\LobsLEdd$, but the effect is limited and does not by itself fully explain the observed downward offsets from the canonical $R-L$ relation. 
The remaining discrepancy likely arises from our assumptions regarding the BLR density distribution. In accretion disk models \citep[e.g.][]{Shakura&Sunyaev1973, Abramowicz1988, Chen&Wang2004, Czerny2019}, the disk gas density depends on both BH mass and accretion rate. If the BLR gas is coupled to the accretion flow, these scalings naturally suggest higher BLR gas densities in lower-mass and higher-accretion-rate systems. We explore the impact of this BLR density enhancement in more detail in \S~\ref{sec:var_blr_den_mbh} and \S~\ref{sec:var_blr_den_lledd}.

The BLR gas density is typically modeled with hydrogen densities \nH\ spanning $10^8$ to $10^{12}$ cm$^{-3}$ in the LOC model. At densities below $10^{8}$\,cm$^{-3}$, many permitted lines become inefficient in emission under typical BLR ionizing fluxes, and broadened forbidden lines would be expected to appear, which are generally not observed in classical broad-line spectra; whereas at densities above $\sim10^{14}$ cm$^{-3}$, the majority of the gas clouds are thermalized and continuum-dominated \citep{Baldwin+1995, Ferguson+1997, Korista&Goad_2000}. While the physical parameters adopted in the LOC framework are often fixed to the standard ranges, BLR emissivity may be a multi-parameter phenomenon, governed by the interplay between disk geometry, gas density, and the hardness of the ionizing continuum. Thus, we will examine how variations in these parameters affect the emissivity-weighted BLR distance.

\subsubsection{Dependence on BLR cloud density with BH mass}\label{sec:var_blr_den_mbh}

\begin{figure*}
    \includegraphics[width=\linewidth]{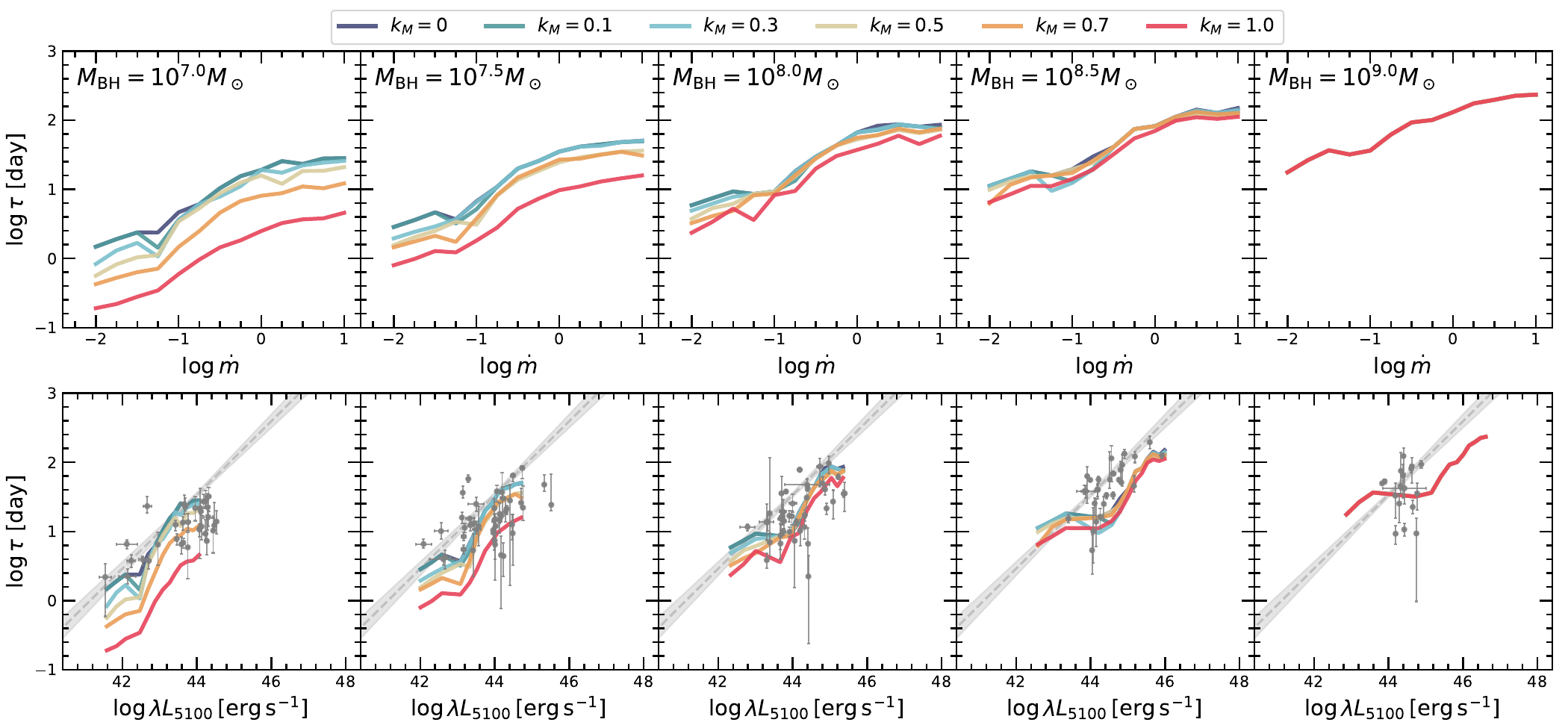}
    \caption{The predicted \hbeta\ emissivity-weighted BLR radius, used as a proxy for the observed time lag,  as a function of accretion rate and optical luminosity for SMBHs with $10^7$ to $10^9 \,\msun$, assuming a BLR density enhancement parameter $k_{M}$ that depends on BH mass. Curves are color-coded by the density enhancement parameter $k_{M}$.
    Top panels: Emissivity-weighted BLR radius $\tau=R_{\rm BLR}/c$ as a function of dimensionless accretion rate \dotm. Bottom panels: The model-predicted $R-L$ relation overlaid on RM measurements (gray points). The gray dashed line shows the canonical $R-L$ relation \citep{Bentz2013}. 
    }\label{fig:tau_vary_nH_Mbh}
\end{figure*}

\begin{table}
\centering
\begin{tabular}{cccccc}
\hline\hline
\multicolumn{6}{c}{Median $\log L_{\rm H\beta}$ [$\ergs$]} \\ \hline
$\Gamma$ & \multicolumn{5}{c}{BLR density enhancement factor $k_{M}$ } \\
\cline{2-6} & $0.1$ & $0.3$ & $0.5$ & $0.7$ & $1.0$  \\
\hline
$-1.6$ & 42.01 & 42.03 & 42.06 & 42.08 & 42.12 \\
$\mathbf{-1.5}$ & \textbf{42.50} & \textbf{42.50} & \textbf{42.50} & \textbf{42.51} & \textbf{42.54} \\
$-1.4$ & 42.89 & 42.91 & 42.94 & 42.95 & 42.97 \\
$-1.3$ & 43.34 & 43.35 & 43.38 & 43.37 & 43.41 \\
$-1.2$ & 43.79 & 43.80 & 43.83 & 43.84 & 43.86 \\
$-1.1$ & 44.24 & 44.25 & 44.29 & 44.29 & 44.32 \\
$-1.0$ & 44.75 & 44.79 & 44.80 & 44.81 & 44.83 \\
\hline
\hline
\end{tabular}
\caption{Median H$\beta$ line luminosity ($\log L_{\rm H\beta}$) predicted by the LOC model with different BLR density enhancement $k_{M}$ as a function of BH mass. Boldface values indicate the best-matching $\Gamma$ for each $k_{M}$.}
\label{tab:median_lhb_gamma_nvar_mbh}
\end{table}

In the SSD theory, the local disk gas density in the outer gas-pressure-dominated region decreases with increasing BH mass at fixed accretion rate, with $\rho \propto \Mbh^{-7/10}$ \citep{Shakura&Sunyaev1973, Kato+2008}. A similar trend is present in slim-disk solutions, where lower-mass BHs have denser accretion flows than higher-mass systems \citep{Chen&Wang2004, Czerny2019}. If we assume that the BLR gas inherits the density scaling of the accretion flow, then lower-mass systems should have systematically denser line-emitting gas. Since photoionized line emissivity is density dependent, this effect can shift the effective BLR radius and affect the normalization of the $R-L$ relation. To explore this effect, we parameterize the BH-mass-dependent density enhancement as
\begin{equation}
    \Delta \log\nH = - k_{M}\times\log\Delta\Mbh
\end{equation}
and adjust the density integration range in Equation~\ref{eq:loc_integral} accordingly. For example, with $k_{M}=0.5$, the fiducial density range $\log\nH=8-12$ remains unchanged at $\log\Mbh=9$, but shifts to $\log\nH=8.5-12.5$ at $\log\Mbh=8$, and to $\log\nH=9-13$ at $\log\Mbh=7$. 
We also tested variations in the radial index $\Gamma$ and found that the density enhancement has only a secondary impact on the total line luminosity compared to $\Gamma$, as listed in Table~\ref{tab:median_lhb_gamma_nvar_mbh}. Among the tested values, $\Gamma=-1.5$ provides the optimal match to the observed \hbeta\ luminosity normalization (e.g., lower-left panel of Figure~\ref{fig:Lline_hbeta_gamma}). We therefore adopt a constant value of $\Gamma=-1.5$ throughout the BH-mass-dependent density-enhancement models.

The effects of BH-mass-dependent density enhancement are shown in Figure~\ref{fig:tau_vary_nH_Mbh}.  The top panels show the emissivity-weighted lag $\tau$ as a function of accretion rate for different \Mbh\ and density enhancements $k_{M}$, while the bottom panels map these predictions into the observational $R-L$ plane.
Increasing $k_M$ mainly affects the lower-mass systems, where the BLR lag is progressively reduced relative to the baseline model with constant density ($k_{M}=0$). The effect is modest at high BH mass, but becomes significant for $\Mbh\lesssim10^8\,\msun$, where the denser BLR gas shifts the emissivity-weighted \hbeta\ emission to smaller radii. 
In the $R-L$ plane, the BH-mass-dependent density enhancement pulls the model sequences downward, and therefore helps reduce the discrepancy for high-accretion-rate objects whose observed RM lags lie below the canonical $R-L$ relation. However, this improvement is not uniform across the sample. In particular, stronger density enhancement also lowers the predicted radii for low-accretion-rate objects, which were already reasonably consistent with the fiducial LOC model, thereby degrading the agreement in that part of parameter space. 
Thus, a pure BH-mass-dependent density prescription does not provide an overall improvement relative to the fiducial model, even though it captures the qualitative direction needed to explain the shortened lags of high-\LLEdd\ sources. We therefore do not interpret the BH-mass-dependent scaling as a unique or globally preferred solution. Instead, it serves as a physically motivated test case for how the BLR size would respond if the BLR gas density were coupled to the density scaling of a standard accretion disk. Motivated by the approximate mass dependence of the local SSD density, we adopt $k_M=0.7$ as an illustrative case in the following comparison. This choice demonstrates that density enhancement can substantially reduce the predicted \hbeta\ emissivity-weighted radius and help explain the offsets of high-accretion-rate AGNs, while also indicating that additional dependence on accretion rate is likely required to preserve consistency with the low-accretion-rate population.

\subsubsection{Dependence on BLR cloud density with accretion rate}\label{sec:var_blr_den_lledd}

\begin{figure*}
    \includegraphics[width=\linewidth]{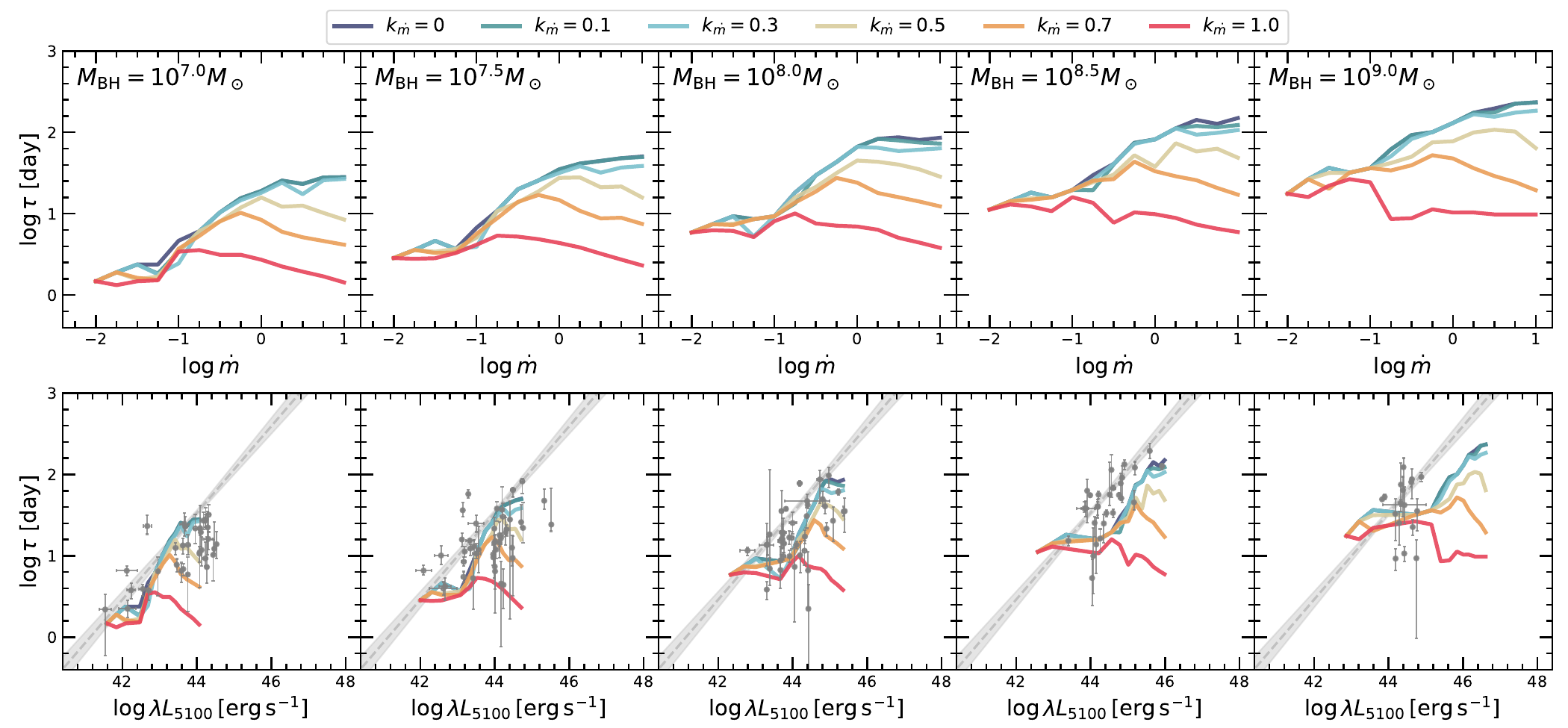}
    \caption{Same as Figure~\ref{fig:tau_vary_nH_Mbh}, but with the BLR density enhancement $k_{\dotm}$ parameterized by the dimensionless accretion rate \dotm. 
    } \label{fig:tau_vary_nH_dotm}
\end{figure*}

\begin{table}
\centering
\begin{tabular}{cccccc}
\hline\hline
\multicolumn{6}{c}{Median $\log L_{\rm H\beta}$ [$\ergs$]} \\ \hline
$\Gamma$ & \multicolumn{5}{c}{BLR density enhancement factor $k_{\dot{m}}$ } \\
\cline{2-6} & $0.1$ & $0.3$ & $0.5$ & $0.7$ & $1.0$  \\
\hline
$-1.6$ & 42.02 & 42.04 & 42.08 & 42.10 & 42.16 \\
$\mathbf{-1.5}$ & \textbf{42.49} & \textbf{42.50} & \textbf{42.51} & \textbf{42.52} & \textbf{42.56} \\
$-1.4$ & 42.89 & 42.92 & 42.94 & 42.95 & 42.99 \\
$-1.3$ & 43.34 & 43.36 & 43.38 & 43.40 & 43.42 \\
$-1.2$ & 43.79 & 43.81 & 43.83 & 43.84 & 43.87 \\
$-1.1$ & 44.24 & 44.27 & 44.32 & 44.30 & 44.31 \\
$-1.0$ & 44.75 & 44.79 & 44.82 & 44.83 & 44.81 \\
\hline
\hline
\end{tabular}
\caption{Same as Table~\ref{tab:median_lhb_gamma_nvar_mbh} but with different BLR density enhancement $k_{\dot{m}}$ as a function of accretion rate.}
\label{tab:median_lhb_gamma_nvar_lledd}
\end{table}

Recent multi-scale theoretical simulations in \citet{Hopkins+2024, Hopkins2025_FIRE} predicted that the BLR gas forms a clumpy thick-disk geometry. In their models, the BLR gas densities span approximately $\sim 10^9 - 10^{13}$ cm$^{-3}$, with gas-mass-weighted mean densities around $10^{11}$ cm$^{-3}$, comparable to traditional photoionization BLR values. Importantly, these simulations indicate that increasing accretion rate leads to higher gas densities in the radial range corresponding to typical BLR locations. Direct observational constraints on this trend remain limited, as BLR gas densities cannot be measured independently without relying on photoionization modeling. Nevertheless, \citet{Negrete+2012} analyzed two high-accretion-rate AGNs using a single-cloud photoionization model and inferred high BLR densities of $\nH\sim10^{13}{\rm cm^{-3}}$, broadly consistent with the expectation that BLR gas density may increase with accretion rate.
Additionally, \citet{Panda+2019_WarmCorona, Panda+2019_MainSeq} modeled the quasar main sequence and found that the strongest optical \FeII\ emitters, which are typically associated with high-\LLEdd\ sources, require dense, high-column line-emitting gas in addition to variations in Eddington ratio, metallicity, and orientation, suggesting that rapidly accreting systems tend to host denser line-emitting gas. 
A similar density increase is expected in the accretion flow from accretion disk theory. In the outer $p_{\rm gas}$-dominant region of an SSD, the local gas density increases with accretion rate as $\rho\propto\dotm^{11/20}$ \citep{Shakura&Sunyaev1973, Kato+2008}. Slim-disk solutions likewise predict that the local disk density depends on the accretion rate at the outer region, although the exact scaling differs from that in the thin-disk case and demonstrates a complicated trend \citep{Chen&Wang2004, Czerny2019, Feng+2019}. 

Motivated by these considerations, we extend our fiducial BLR setup by allowing the characteristic density range that contributes to the emissivity-weighted integration to increase with the accretion rate in Equation~\ref{eq:loc_integral}. Specifically, we parameterize the density enhancement as
\begin{equation}
    \Delta \log\nH =  k_{\dotm}\times\log\Delta\dotm
\end{equation}
For example, with $k_{\dotm}=0.5$, the characteristic BLR gas density range in the LOC model shifts from the baseline $\log\nH=8-12$ at $\log\dotm=-2$, to $\log\nH=8.5-12.5$ at $\log\dotm=-1$, and to $\log\nH=9-13$ at $\log\dotm=0$. The radial index is fixed at $\Gamma=-1.5$ to match the overall \hbeta\ luminosity normalization, as shown in the lower-right panel of Figure~\ref{fig:Lline_hbeta_gamma} and summarized in Table~\ref{tab:median_lhb_gamma_nvar_lledd}.

Figure~\ref{fig:tau_vary_nH_dotm} illustrates the impact of density-dependent BLR structure on the predicted \hbeta\ lag.
At low accretion rates ($\log\dotm\lesssim-1$), variations in the assumed cloud density only have a limited effect on the predicted lag. In this regime, the SED is relatively hard, and the line emissivity is less sensitive to density compared to the near-Eddington regime with softer SEDs. At higher accretion rates, the behavior changes significantly. In the absence of any density enhancement ($k_{\dotm}=0$, dark blue curves), the BLR distance initially increases and then mildly increases or flattens with increasing \dotm, reflecting the combined effects of disk thickening and reduced ionizing flux at BLR latitudes. When density enhancement is included ($k_{\dotm}>0$), the lag at high accretion rates is progressively suppressed. This trend arises because the softer ionizing SED is more efficiently reprocessed by higher-density gas, which preferentially emits at smaller radii and shifts the emissivity-weighted BLR inward. 
The fiducial model with constant density range ($k_{\dotm}=0$) systematically overpredicts the BLR size at high accretion rate, while these accretion-rate-dependent models with density enhancement ($k_{\dotm}\gtrsim0.5$) exhibit a pronounced turnover in the $R-L$ plane and simultaneously reproduce the offsets of the observed high-accretion population. Extremely large density enhancements tend to over-suppress the lag, but a moderate enhancement, such as $k_{\dotm}=0.5$  (beige curve), provides a plausible description of the observed deviations from the canonical $R-L$ relation while preserving the low-\dotm\ behavior relatively well.

\begin{figure*}
    \includegraphics[width=\linewidth]{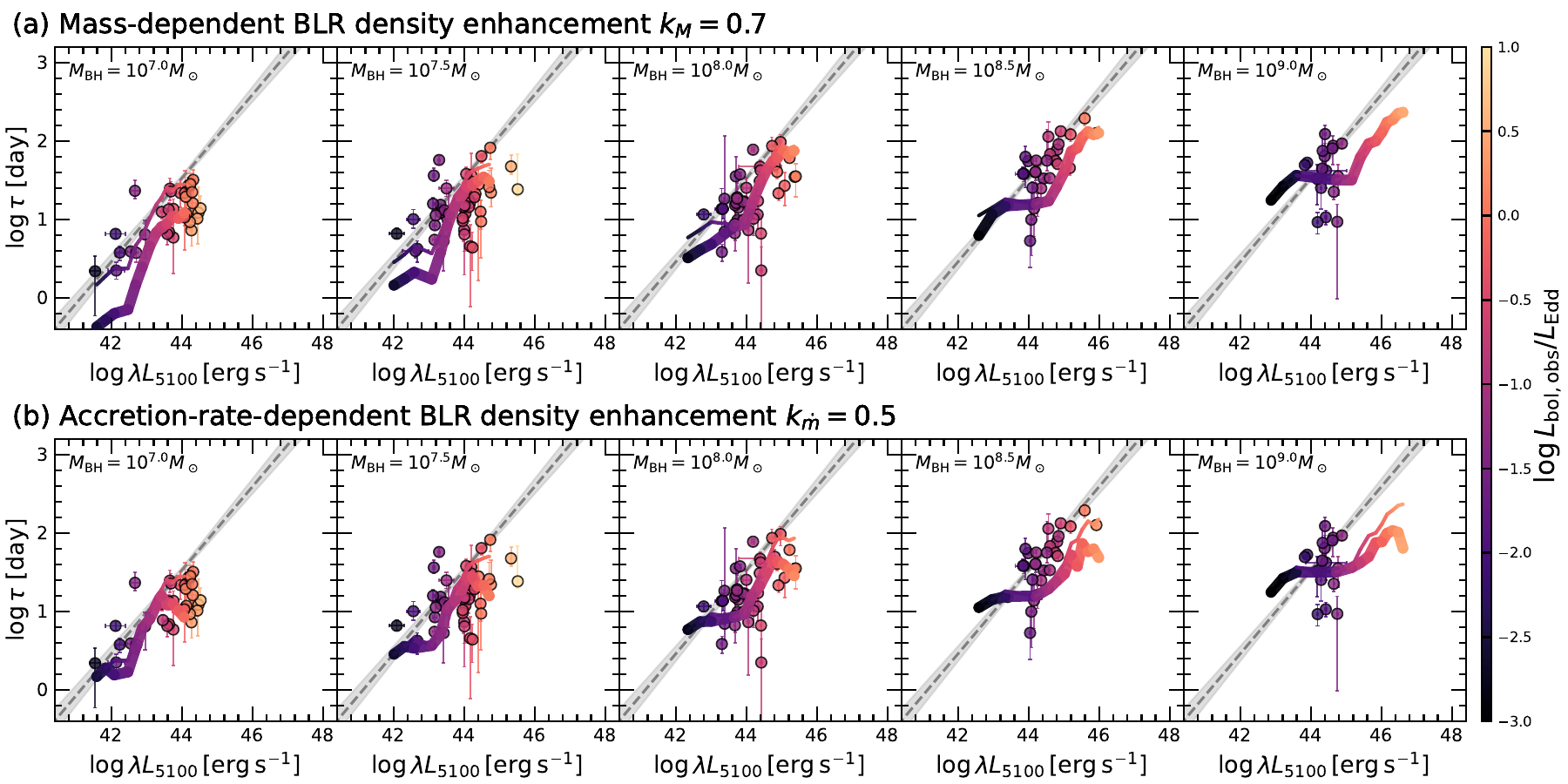}
    \caption{Predicted \hbeta\ BLR size–luminosity relation for models with density enhancement tied to BH mass and accretion rate. The enhanced-density model predictions are shown as thick colored curves, while the corresponding fiducial LOC model predictions are shown as narrow curves. Both the data and model predictions are color-coded by observed Eddington ratio \LobsLEdd. The gray dashed lines mark the canonical $R-L$ relation \citep{Bentz2013}. Top (a): model predictions with a BH-mass-dependent BLR density enhancement ($k_M=0.7$). Bottom (b): model predictions with an accretion-rate-dependent BLR density enhancement ($k_{\dotm}=0.5$).} 
    \label{fig:R-L_offsets}
\end{figure*}

We summarize the effects of BLR density enhancement on the $R-L$ relation in Figure~\ref{fig:R-L_offsets}. Both the BH-mass-dependent and accretion-rate-dependent prescriptions shift the predicted \hbeta\ lags below the constant-density baseline, but they modify the $R-L$ relation in different ways. 
The BH-mass-dependent model primarily changes the normalization as a function of BH mass, producing the strongest lag reduction in lower-\Mbh\ systems; this helps explain some high-accretion objects but does not improve the global agreement because it also lowers the predicted radii for low-accretion systems that are already consistent with the fiducial model. By contrast, the accretion-rate-dependent model mainly acts along the \LLEdd\ sequence, producing a stronger turnover at high accretion rate and more naturally matching the observed downward offsets of high-\LLEdd\ AGNs. A decisive test requires RM samples that densely populate the two-dimensional \Mbh--\LLEdd\ plane, especially high-\Mbh, high-\LLEdd\ AGNs, where the two models predict the most different behavior. 
Taken together, these results suggest that both BH mass and accretion rate may regulate the characteristic BLR density, and that the observed scatter and systematic offsets from the canonical $R-L$ relation are likely driven by a combination of these effects rather than by self-shadowing alone. Phenomenologically, however, the accretion-rate-dependent density enhancement is favored, especially for low-\Mbh\ systems, because it reproduces the shortened lags observed in high-accretion-rate AGNs \citep{Du+2016, Du2018_SEAMBH, Hu2021_SEAMBH} without degrading the agreement for low-\dotm\ objects as strongly as the purely BH-mass-dependent model.

\subsubsection{Dependence on BLR opening angle}

Recent dynamical modeling and interferometric results favor a geometrically thick distribution of BLR with an opening angle $\theta_{\rm BLR}\sim30^\circ$ to $60^\circ$ \citep{Williams+2018, Gravity_Abuter+2024, Gravity+2024, ZStone+2025}. These studies suggest that BLR clouds occupy an equatorial configuration with substantial vertical extent, rather than forming either a spherical distribution or a geometrically thin disk.

\begin{figure*}
    \includegraphics[width=\linewidth]{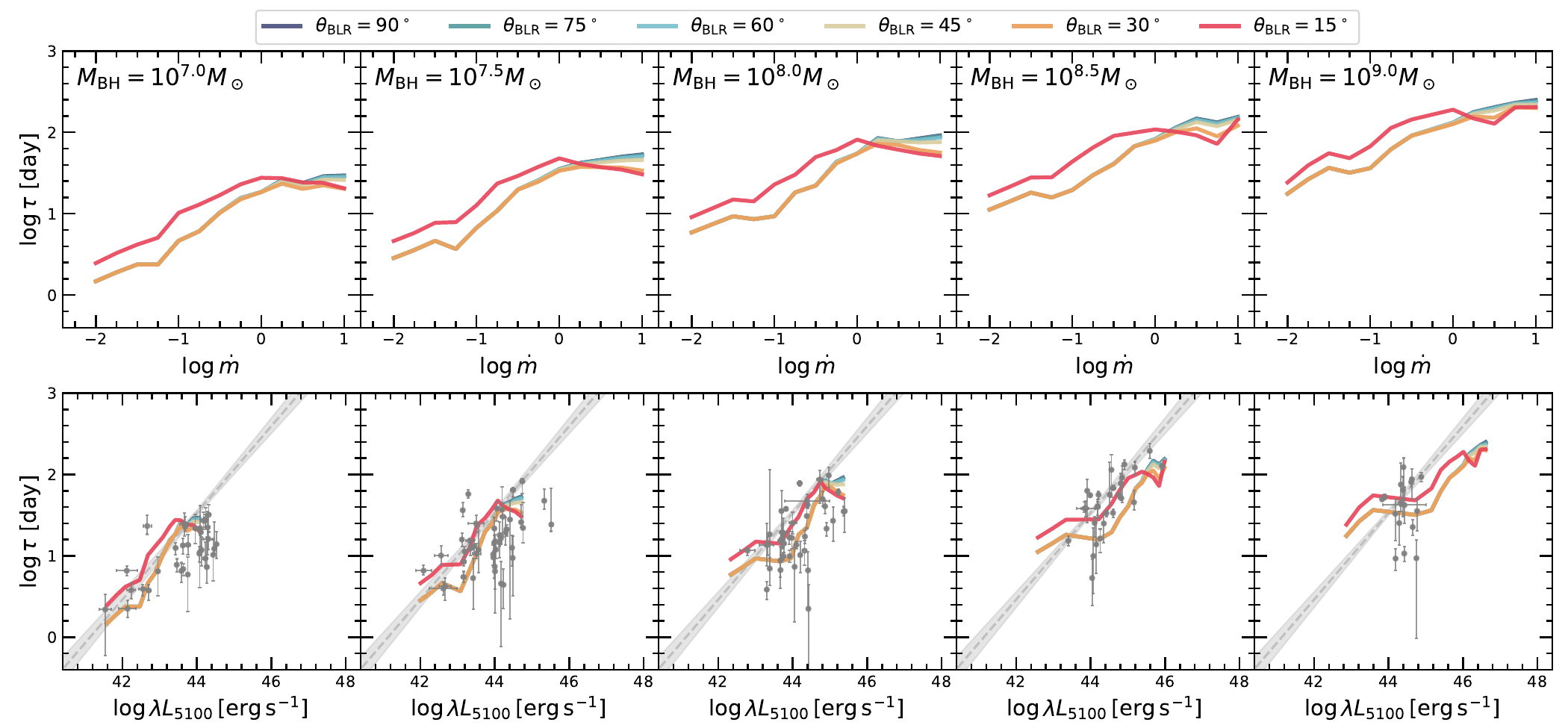}
    \caption{Same as Figure~\ref{fig:tau_vary_nH_Mbh}, but with different BLR opening angle $\theta_{\rm BLR}$ and radial index $\Gamma$.}\label{fig:tau_vary_open}
\end{figure*}

In our anisotropic-illumination model, the adopted BLR opening angle is partially degenerate with the effective covering of the ionizing source, because changing $\theta_{\rm BLR}$ changes the solid angle occupied by line-emitting gas and therefore changes the relative contribution of clouds at different latitudes to the total line luminosity. This geometric effect is distinct from the global covering factor, $CF$, which describes the fraction of the ionizing continuum intercepted by BLR clouds and primarily sets the overall normalization of the line luminosity \citep{Korista&Goad_2000}. We fix the global covering factor to $CF=50\%$ and vary only the angular extent of the BLR in our computation.

Within the LOC framework, \citet{Korista&Goad_2000} showed that the BLR covering factor ($CF$) is anti-correlated with the radial distribution index $\Gamma$. In particular, a larger covering factor requires a steeper radial weighting, i.e., a more negative $\Gamma$, to avoid overproducing line emission from large radii. 
A similar coupling between BLR opening angle and the radial weighting index appears in our model, as summarized in Table~\ref{tab:median_lhb_cf_gamma}. For each adopted BLR opening angle, $\Gamma$ must be recalibrated to reproduce the observed median \hbeta\ luminosity; therefore, for each $\theta_{\rm BLR}$, we adopt the value of $\Gamma$ that best matches the observed median \hbeta\ luminosity, as indicated by the boldfaced entries in Table~\ref{tab:median_lhb_cf_gamma}. We then use these calibrated values of $\Gamma$ to compute the predicted \hbeta\ emissivity-weighted BLR radius for different BLR opening angles $\theta_{\rm BLR}$.

Figure~\ref{fig:tau_vary_open} illustrates how the predicted \hbeta\ lag depends jointly on BH mass, accretion rate, and BLR angular extent. At fixed \Mbh, the predictions for different opening angles broadly follow the behavior of our fiducial model: the \hbeta\ lag increases with increasing $\dot{m}$ and exhibits a mild turnover at high $\dot{m}$ due to disk self-shadowing. 
The dependence on BLR opening angle is more complex, because variations in $\theta_{\rm BLR}$ are coupled to changes in the radial weighting index $\Gamma$, thus, the curves in Figure~\ref{fig:tau_vary_open} should be interpreted as the combined response of the BLR geometry and the recalibrated cloud distribution, rather than as the effect of opening angle alone.
For large BLR opening angles ($\theta_{\rm BLR}\geq45^\circ$), a substantial fraction of the line emission arises from high-altitude clouds, which dominate the emissivity-weighted BLR radius. 
In the low-accretion-rate regime ($\log\dot{m}<0$), the SSD remains geometrically thin, so self-shadowing is negligible and the BLR opening angle has only a weak effect on the effective \hbeta\ radius. As the accretion rate increases, the puffed-up slim disk begins to shield lower-latitude clouds from the ionizing continuum, leading to a slightly stronger shadowing effect for smaller $\theta_{\rm BLR}$. 
For small BLR opening angles ($\theta_{\rm BLR}\leq30^\circ$), many of the emitting clouds lie within the angular region most affected by disk self-shadowing, especially for high-accretion rate cases($\log\dotm\geq0$). Consequently, the predicted \hbeta\ lag shows a more pronounced turnover on the right. 
However, because smaller opening angles also correspond to a smaller effective emitting solid angle, these geometrically restricted BLR configurations would underproduce the total \hbeta\ luminosity unless the cloud distribution is adjusted. The adopted $\Gamma$ therefore partly compensates for the reduced illuminated cloud volume by assigning greater weight to larger radii, which increases the emissivity-weighted BLR radius and moderates the expected reduction in the \hbeta\ lag.
The dependence of the \hbeta\ lag on $\theta_{\rm BLR}$ is therefore not monotonic in our model, but instead reflects the combined effects of BLR geometry, anisotropic illumination, and the luminosity-calibrated radial cloud distribution. Because $CF$ and $\theta_{\rm BLR}$ both control the amount of emitting gas exposed to the ionizing continuum, the present calculations cannot uniquely separate a change in global covering factor from a change in BLR angular extent.

Theoretical models also suggest that the BLR opening angle is not fixed, but may evolve with accretion state. In particular, the FRADO (Failed Radiatively Accelerated Dusty Outflow) model naturally gives rise to a flattened BLR with substantial vertical extent \citep{Czerny&Hryniewicz_2011_frado, Czerny+2017_FRADO}. Disk-wind models similarly suggest that the launching angle and angular extent of the outflow are sensitive to luminosity and radiation pressure, implying that the vertical distribution of BLR gas may vary systematically with Eddington ratio \citep{Giustini&Proga2019}. These considerations suggest that a more self-consistent treatment should couple the angle-dependent slim-disk radiation field to the physical origin and dynamics of BLR gas, including cloud formation and wind launching. In this context, combining a physically motivated BLR formation scenario such as FRADO with self-shadowed photoionization calculations would be a particularly interesting direction for future work.
Radiation magnetohydrodynamic (MHD) simulations that include disk vertical structure, radiation pressure, and line-driven or dusty wind launching would be particularly valuable for determining whether the BLR angular extent evolves with accretion rate and how this evolution affects the observed \hbeta\ $R-L$ relation.

\subsection{Comparison with MHD simulations}\label{sec:MHD_sims}

Three-dimensional radiation MHD simulations have substantially advanced our understanding of BH accretion flows over the last decade \citep[e.g.][]{McKinney+2014_MHD, Jiang+2019_SubEdd, Jiang+2019_SuperEdd, Hopkins+2024, Hopkins2025_FIRE}. Simulations of super- and sub-Eddington regimes consistently demonstrate the formation of geometrically thick, radiation-pressure-dominated disks with wide polar funnels in inner regions \citep{Jiang+2019_SubEdd, Jiang+2019_SuperEdd}. Their computational domain typically resolves radii $\lesssim 10^3\Rg$, whereas the BLR typically extends from $\sim 10^3$ to $10^5\,\Rg$. On larger scales, galaxy-scale simulations show that anisotropic radiation pressure, AGN-driven winds, and gravitational instabilities naturally produce multiphase, clumpy gas in the central parsecs, with densities, covering fractions, and dynamical states broadly consistent with BLR environments \citep{Hopkins+2024, Hopkins2025_FIRE}. These simulations provide important physical insights into the gas distribution and kinematics in the immediate vicinity of SMBHs; however, they cannot yet be directly connected to BLR emissivities as they do not fully resolve the EUV and soft X-ray ionizing continuum most relevant for BLR photoionization. As a result, detailed emission-line properties cannot be predicted directly from existing simulations and require separate photoionization modeling.

Our framework represents a simplified disk-BLR photoionization geometry and inherently excludes the complex gas dynamics self-consistently captured in MHD simulations. We also do not model the magnetically driven winds or outflows that play critical roles in angular momentum transport and vertical energy redistribution \citep[e.g.,][]{Blandford&Payne1982, McKinney+2014_MHD, Sadowski+2014}.
Instead, we adopt a parameterized disk structure designed to reproduce the large-scale disk thickening predicted by these simulations. We then couple these SEDs to a standard photoionization framework to evaluate their impact on BLR emissivity and inferred radii. Our assumptions regarding gas density and radial scales are broadly consistent with observations, though we do not attempt to model their dynamical evolution. Accordingly, our goal is not to reproduce the dynamical complexity of MHD simulations but to provide a phenomenological bridge between physically motivated accretion disk models and observable BLR properties. 

Future MHD simulations that incorporate frequency-dependent radiation transport and extend to larger radial domains may eventually provide angle-resolved ionizing flux distributions, enabling fully self-consistent coupling between accretion flow dynamics and BLR photoionization. Conversely, observational constraints on gas properties in the immediate AGN environment, such as BLR sizes and emission-line flux, can provide critical insights into the accretion physics and radiation anisotropy in MHD simulations.

\subsection{Limitations and future work}

Compared with our previous work \citepalias{WuQ+2025}, which directly employed the three-component AGN SED model {\tt qsosed}, this study extends the accretion-rate range to near- and super-Eddington regimes.  Observational comparisons between {\tt qsosed} predictions and stacked AGN SEDs \citep{Kynoch_etal2023, Mitchell_etal2023, Hagen+2024} indicate that the model performs well for black hole masses in the range of $10^{7.5}-10^9\msun$, but tends to underestimate the X-ray flux at lower \Mbh\ and exhibits discrepancies in the optical/UV continuum for higher-\Mbh\ systems. 
These SEDs should be interpreted with caution because they are calibrated on observationally inferred BH masses, often based on single-epoch virial estimates with substantial systematic uncertainties \citep{Shen&Liu2012, Shen+2024, Buchner+2026}. Such uncertainties may place some objects in artificially high-mass bins, contributing to the mismatch between SED models and stacked observations \citep{Mitchell_etal2023} and partly explaining the slight BLR-size discrepancy in our $\geq10^{8.5}\,\msun$ models.
Our previous work adopted a fixed photoionization parameter to reproduce the overall $R-L$ trend in the sub-Eddington regime, but predicted an inconsistent trend with observations as the accretion rate approached the Eddington limit. To address this discrepancy, we implement an analytic slim-disk solution to derive a more physically motivated SED in the EUV and soft X-ray regime, with an improved treatment of radiation anisotropies at high accretion rates.
However, the physical origin of the warm Comptonization component and its coupling to the hot corona remain poorly understood \citep{Petrucci_etal2018}. These uncertainties cannot be resolved simply by adopting an analytic slim-disk prescription. A self-consistent model that connects the warm and hot Comptonization regions within a unified physical framework would enable more robust comparisons between theoretical SEDs and observations, and provide a more comprehensive description of the accretion flow near the central engine.

Although our LOC framework provides a flexible and phenomenological description of BLR emission by integrating line emissivity over broad distributions of gas density and radial distance, it does not specify the dynamics, physical origin, confinement mechanism, or stability of the clouds \citep{Czerny&Hryniewicz_2011_frado, Czerny+2017_FRADO, Baskin2014_RPC}. The density and radial weighting functions are empirically adopted rather than derived from the underlying accretion-flow dynamics. 
While this LOC approach successfully reproduces many observed emission-line properties, interpretations based on LOC prescriptions remain sensitive to the assumed parameter distributions. An alternative framework for BLR cloud structure is the radiation-pressure confinement (RPC) model, in which the gas density is regulated by the local incident radiation field rather than prescribed independently \citep{Stern+2014_RPC, Baskin2014_RPC}. While RPC is physically well motivated, we do not adopt it here because our primary goal is to isolate the effects of anisotropic illumination on the emissivity-weighted BLR radius while allowing the cloud density distribution to vary in a controlled manner. In the RPC framework, the gas ionization structure and line emission are directly coupled to the BLR slab distance, making it difficult to disentangle the specific roles of self-shadowing and density enhancement within a global two-dimensional parameter space. We therefore retain the LOC framework, which provides a more flexible parameterization for exploring the effects of self-shadowing and for reproducing the properties of high-ionization UV emission lines. A self-consistent extension combining RPC with angle-dependent slim-disk illumination would be a valuable direction for future work.

We also caution that our predicted BLR radii are \hbeta\ emissivity-weighted radii rather than direct RM lags. Reverberation mapping measures the variable line response to continuum fluctuations, whereas our calculation characterizes the spatial distribution of the time-averaged \hbeta\ emissivity. These two quantities may not be identical, because the measured lag depends on the transfer function, continuum variability, BLR geometry, and local line response \citep{Goad+1993, Korista&Goad2014, Goad&Korista2015, Li&Wang2025}. In photoionization-based interpretations, responsivity-weighted radii are often introduced to describe the variable component of the line emission, since regions that dominate the total \hbeta\ luminosity do not necessarily dominate the responding component \citep{Korista&Goad2014, Li&Wang2025}. 
However, responsivity-weighted quantities are model-dependent and rely on assumptions about the BLR structure, varying ionizing continuum, and time-dependent line-response physics. We therefore use the emissivity-weighted radius as a first-order proxy for the characteristic BLR scale, which allows us to isolate the systematic effects of accretion-rate-dependent SED anisotropy and density enhancement on the predicted \hbeta\ emitting region. 

Moreover, as discussed in \S\ref{sec:MHD_sims}, multi-scale simulations suggest that the basic gas properties in the central parsec broadly agree with those assumed in our BLR models, but key physical processes such as outflows, dynamical evolution, and magnetic fields \citep[e.g.,][]{Blandford&Payne1982, McKinney+2014_MHD, Sadowski+2014, Jiang+2019_SubEdd, Jiang+2019_SuperEdd} are not explicitly included in this work. One natural extension of this research involves the development of a frequency-dependent radiative transfer with photoionization modeling to move beyond the parametric assumptions of current models by self-consistently tracking the propagation of the ionizing continuum through a dynamical gasous region. By exploring a wider parameter space of accretion rates and BH masses with fine grids, future studies can delineate the interplay between accretion flows and the line-emitting regions and the growth of SMBHs across cosmic time.

\section{Conclusions}\label{sec:conclusion}

Recent RM studies have shown that high-accretion-rate AGNs systematically lie below the canonical BLR $R-L$ relation, exhibiting shorter lags than predicted at fixed optical luminosity. In this work, we investigate whether slim-disk self-shadowing, together with variations in BLR gas density and geometry, can account for these deviations. We construct a two-dimensional BLR grid illuminated by physically motivated, angle-dependent SEDs, enabling us to evaluate how anisotropic radiation and BLR cloud properties shape the effective line-emitting region.
Our main findings are summarized as follows:
\begin{enumerate}
    \item{
    In the sub-Eddington regime, the disk remains well described by the SSD solution, and the predicted BLR radius follows the canonical $R\propto L^{1/2}$ scaling. As the accretion rate approaches and exceeds the Eddington limit, slim-disk effects modify the disk temperature profile and produce a softened, anisotropic SED, thereby changing the ionizing photon field seen by the BLR (Figure~\ref{fig:agnsed_vs_slim}). }
    \item{
    Geometric thickening of the inner slim disk produces self-shadowing that preferentially suppresses the ionizing flux received by BLR clouds at low latitudes (Figure~\ref{fig:sed_self_shadow}). This anisotropic illumination shifts the emissivity-weighted \hbeta\ emission inward and reduces the effective BLR radius relative to the unshielded case (Figure~\ref{fig:loc_shield_vs_unshield}). The suppression becomes stronger with increasing accretion rate, naturally flattening the predicted BLR size at high \LLEdd. The self-shadowed models reproduce the overall observed trend in the \hbeta\ $R-L$ plane reasonably well. However, self-shadowing alone does not fully account for the magnitude of the observed offset, particularly in lower-mass, high-accretion-rate systems, indicating that additional changes in BLR gas properties are likely required.}
    \item{
    Motivated by accretion-disk density scalings, we explored the possibility that the BLR gas density increases toward lower \Mbh\ if the BLR is physically coupled to the accretion flow. A BH-mass-dependent density enhancement shifts the \hbeta-emitting region to smaller radii, with the strongest effect in low-\Mbh\ systems (Figure~\ref{fig:tau_vary_nH_Mbh}). This prescription can reduce the predicted lag offsets for low-\Mbh, high-\LLEdd\ AGNs, but it does not uniformly improve the agreement with RM measurements because it also lowers the predicted BLR radii for low-accretion-rate objects that are already reasonably consistent with the fiducial LOC model. 
    }
    \item{
    We also examined models in which the characteristic BLR density increases with accretion rate. In this case, higher-density BLR gas at large \dotm\ shifts the effective \hbeta-emitting region inward and introduces a pronounced turnover in the predicted $R-L$ plane (Figure~\ref{fig:tau_vary_nH_dotm}). Compared with the BH-mass-dependent prescription, the accretion-rate-dependent model acts more selectively in the region where the observed offsets are largest, while preserving the low-\dotm\ behavior more effectively. A moderate enhancement with $k_{\dotm}=0.5$ provides a plausible match to the shortened lags observed in high-accretion-rate AGNs, making accretion-rate-dependent BLR density evolution the phenomenologically preferred explanation in our model.
    }
    \item{
    Variations in the assumed BLR opening angle also affect the predicted \hbeta\ radius by changing the relative contribution of clouds at different latitudes to the total line emission (Figure~\ref{fig:tau_vary_open}). However, because the BLR angular extent is partially degenerate with the effective covering of the ionizing source and with the LOC radial weighting parameter $\Gamma$, its impact is difficult to isolate. Compared with BLR density evolution, changes in opening angle generally produce a more moderate effect on the emissivity-weighted BLR radius, suggesting that geometry alone is unlikely to explain the largest offsets from the canonical $R-L$ relation. }
\end{enumerate}

Overall, our results suggest that the deviations from the canonical $R-L$ relation in high-accretion-rate AGNs can be understood as a natural consequence of changes in accretion-disk geometry together with systematic variations in BLR physical conditions (Figure~\ref{fig:R-L_offsets}). Although the present framework remains phenomenological, it provides a physically motivated connection between the AGN ionizing SED and BLR observables through photoionization modeling. 
Extending this framework to include more self-consistent BLR formation and confinement scenarios, such as FRADO \citep{Czerny+2017_FRADO} and RPC \citep{Baskin2014_RPC}, time-dependent line responsivity and transfer-function calculations, and coupling to radiation-MHD simulations will be a key step toward a unified picture of the BLR within the broader accretion-flow environment.

\begin{acknowledgments}.
We thank Jian-Min Wang, Yan-Rong Li, Kirk T. Korista, Wei-Min Gu, Yan-Fei Jiang, Bożena Czerny, Zachary Stone, and Lizhong Zhang for helpful suggestions. We are grateful to Gary Ferland and collaborators for the development of CLOUDY and for allowing its free use for the community. This work is partially supported by NSF grant AST-2509424.
\end{acknowledgments}.

\bibliography{refs}{}
\bibliographystyle{aasjournal}

\end{document}